%% file: main.tex
\documentclass{article}[12pt]

\usepackage[utf8]{inputenc} 
\usepackage[T1]{fontenc}    
\usepackage{microtype}
\usepackage{graphicx}
\usepackage{subfigure}
\usepackage{booktabs} 
\usepackage{amsmath,amsfonts,amssymb}
\usepackage[round]{natbib}
\usepackage{url}
\usepackage[capitalize]{cleveref}
\usepackage{multirow}
\usepackage{mathtools}
\usepackage{ marvosym }
\usepackage{tikz}
\usetikzlibrary{bayesnet}
\usepackage{textcomp}
\usepackage{authblk} 
\usepackage{bbm}
\usepackage{array}
\usepackage[margin=1in]{geometry}
\usepackage[title]{appendix}

\makeatletter
\def\@figurecaps{\@ifundefined{tf@pof}{}{\if@filesw
  \immediate\closeout\tf@pof\fi
  \@ifundefined{appendixname}{}{\def\appendixname{}}
  \parindent\z@
  \reset@figtab
  \clearpage}}%
\makeatother

\usepackage{xr}
\newcommand{\bigCI}{\mathrel{\text{\scalebox{1.07}{$\perp\mkern-10mu\perp$}}}}

\newcommand{\R}{\mathbb{R}}
\newcommand{\vnorm}[1]{\left|\left|#1\right|\right|}

\title{Dose-response modeling in high-throughput cancer drug screenings: {{A}}n end-to-end approach}

\author[1,2]{\small Wesley Tansey\thanks{\texttt{wesley.tansey@columbia.edu} (corresponding author)}}
\author[3]{\small Kathy Li}
\author[3]{\small Haoran Zhang}
\author[1,4]{\small Scott W.~Linderman}
\author[2]{\small Raul Rabadan}
\author[1,4,5]{\small David M.~Blei}
\author[1,3]{\small Chris H.~Wiggins}

\affil[1]{\footnotesize Data Science Institute, Columbia University, New York, NY, USA}
\affil[2]{\footnotesize Department of Systems Biology, Columbia University Medical Center, New York, NY, USA}
\affil[3]{\footnotesize Department of Applied Mathematics and Applied Physics, Columbia University, New York, NY, USA}
\affil[4]{\footnotesize Department of Statistics, Columbia University, New York, NY, USA}
\affil[5]{\footnotesize Department of Computer Science, Columbia University, New York, NY, USA}

\begin{document}


\maketitle


\begin{abstract}
{Personalized cancer treatments based on the molecular profile of a patient's tumor are an emerging and exciting class of treatments in oncology. As genomic tumor profiling is becoming more common, targeted treatments to specific molecular alterations are gaining traction. To discover new potential therapeutics that may apply to broad classes of tumors matching some molecular pattern, experimentalists and pharmacologists rely on high-throughput, \textit{in vitro} screens of many compounds against many different cell lines. We propose a hierarchical Bayesian model of how cancer cell lines respond to drugs in these experiments and develop a method for fitting the model to real-world high-throughput screening data. Through a case study, the model is shown to capture nontrivial associations between molecular features and drug response, such as requiring both wild type TP53 and overexpression of MDM2 to be sensitive to Nutlin-3(a). In quantitative benchmarks, the model outperforms a standard approach in biology, with $\approx20\%$ lower predictive error on held out data. When combined with a conditional randomization testing procedure, the model discovers biomarkers of therapeutic response that recapitulate known biology and suggest new avenues for investigation. All code for the paper is publicly available at \url{https://github.com/tansey/deep-dose-response}.}
\end{abstract}

\input{introduction}

\input{experiment}

\input{related_work}


\input{details}



\input{benchmarks}

\input{biomarkers}

\input{discussion}

\section*{Supplementary Material}
The Supplementary material includes the details of our dataset, our preprocessing strategy for molecular features, and the code for all methods described in the paper.

\section*{Acknowledgments}
The authors thank Victor Veitch, Mykola Bordyuh, Antonio Iavarone and Anna Lasorella for many helpful conversations. They also thank Jackson Loper for providing the binary matrix factorization code used in the biomarker testing. WT is supported by the a seed grant from the Data Science Institute of Columbia University and the NIH (U54-CA193313). SWL is supported by the Simons Foundation (SCGB-418011). CHW is supported by the NSF (1305023, 1344668) and NIH (U54-CA193313). DMB is supported by ONR (N00014-17-1-2131, N00014-15-1-2209), NIH (1U01MH115727-01), and DARPA (SD2 FA8750-18-C-0130).
{\it Conflict of Interest}: None declared.

\bibliographystyle{plainnat}
\bibliography{main}


\begin{appendices}

\input{app_plate_layout}

\input{app_features}

\input{app_assumption_checks}

\input{app_cross_contamination}

\input{app_with_cancer_type}

\end{appendices}








\end{document}

%% file: introduction.tex
\section{Introduction}
\label{sec:introduction}
\subsection{High-throughput cancer drug screening}
\label{subsec:introduction:overview}
Genomic sequencing and high-throughput drug screening is becoming cheaper, enabling widespread adoption in both research institutions and hospitals~\citep{muir:etal:2016:sequencing-cost}. Many of these institutions have built datasets that enable scientists to explore potential connections between the molecular profile of a cell (i.e. its DNA and other related biological information) and its phenotypic response to treatment with a certain drug. Specifically in cancer therapeutics, large public datasets have become available with thousands of experiments testing different drugs on different types of cancer cell lines~\citep[e.g.][]{yang:etal:2012:gdsc,barretina:etal:2012:ccle,iorio:etal:2016:gdsc,haverty:etal:2016:reproducible}.

One goal in analyzing these datasets is to build a predictive model of drug response. The model takes in a set of molecular features of a cell line and predicts the expected outcome of using a candidate drug to treat it. The more accurate the predictor, the more it can be trusted to faithfully simulate wet lab results. If a good predictor can be constructed, it can accelerate the discovery of targeted therapies by refining experimentalists' hypotheses much faster than can be done in the lab. Thus, good predictors have high value to biologists. 

We build a generative model of drug response in high-throughput cancer drug screens. The model captures uncertainty inherent in cell line experiments, including measurement error, natural variation in cell growth, and drug response heterogeneity. We take an approximate Bayesian approach to estimating the model parameters and also propose a method for detecting contaminated data. We use the model in a case study analyzing a dataset~\citep{iorio:etal:2016:gdsc} containing hundreds of thousands of experiments. In our study, the model outperforms a state-of-the-art approach for estimating dose-response curves~\citep{vis:etal:2016:sanger-dose-response}. Model predictions also recapitulate known biology involving nonlinear interactions between molecular features and drugs. Approximate dose-response posteriors, in combination with conditional randomization tests~\citep{candes:etal:2018:panning}, reveal both known and potentially novel biomarkers of drug response.

The experimental design used in our case study is similar to many pharmacogenomic profiling experiments. Consequently, we expect the proposed model and insights from our case study to be applicable to other high-throughput cancer drug screening studies. Some studies~\citep{haibe-kains:etal:2013:gdsc-ccle-inconsistency,safikhani:etal:2016:gdsc-ccle-disagreement} have shown results may not transfer between different high-throughput studies, implying the actual model predictions we infer may not perform well on other datasets. Nevertheless, the modeling strategy and techniques we propose for dealing with data-specific challenges in the GDSC can help guide and inform modeling of similar datasets. In the conclusions of the paper, we summarize these insights for practitioners analyzing other datasets, designing new experiments in the lab, or guiding data-gathering policy at hospitals.

%% file: experiment.tex
\subsection{Experimental setup and details}
\label{subsec:introduction:experiment}

Our data come from the Genomics of Drug Sensitivity in Cancer (GDSC)~\citep{yang:etal:2012:gdsc,garnett:etal:2012:gdsc,iorio:etal:2016:gdsc}, a high-throughput screening (HTS) study on therapeutic response in cancer cell lines. The GDSC data comprise the results of testing $1072$ cancer cell lines nearly-combinatorially against $265$ cancer therapeutic drugs, \textit{in vitro}. The experiments were conducted across two separate testing sites: the Wellcome Trust Sanger Institute (``Site 1'') and Massachusetts General Hospital (``Site 2''). Experiments were carried out over the course of multiple years and used one of two assay types, depending on the cell type: suspended (``Assay S'') or adherent (``Assay A''). For a given site and assay type, each experiment was carried out using a series of HTS microwell plates; the layout for each plate is shown in \cref{fig:plate_layout} in the supplement.

A given plate contains hundreds of microwells, where each well is designated as either a negative control, positive control, or treatment. Negative control wells are left unpopulated and untreated, so as to calibrate the base level of machine output if all cells were to die. All other wells are populated with a constant volume of cells; due to natural variation in cell volume, exactly how many cells occupy any given well is unknown. Positive control wells are left untreated so as to measure the base response for drugs that have no effect.\footnote{The terms ``negative'' and ``positive'' control here are used differently from their common usage in biology. However, this is the terminology used by \citet{garnett:etal:2012:gdsc}; we follow their usage.} Finally, a series of drugs are applied to the remaining wells, with each well receiving a different concentration. A single plate is used to test only one specific cell line; all wells in a plate are filled with the specific assay fluid appropriate for the target cell line. Drug concentration ranges were derived based on previous experiments and were delivered at either 5 or 9 dose levels, depending on the testing site. Both testing sites start at the same maximum dose level, which is diluted at a 4-fold rate for the 5 dose schedule and at a 2-fold rate for the 9 dose schedule, resulting in the same final minimum dose. Since the sites use the same minimum and maximum dose, this dilution procedure yields missing data for odd-numbered dose levels in the dataset. Once treated, cells are left to either grow or die for 72 hours.


Cell population size is approximated by a fluorescence assay. A fluorescent compound is added to the wells that is capable of penetrating cells and binding to a protein kept at relatively-constant levels in all living cells. When a cell dies, its structure breaks down and the target protein denatures, leaving no binding agent and resulting in no fluorescence. Robotically-controlled cameras photograph each well and the total luminescence of the image (i.e. pixel intensity count) is used as a relative measure of cell population size.

Luminescence of positive and negative control wells is used to calibrate luminescence of treatment wells. 
Positive control wells provide an estimate for how much light a well with all living cells will emit, while negative control wells provide an estimate for how much light a well with no living cells will emit.
We refer to this negative control measure as the baseline fluorescence bias, since it represents the fluorescence of an empty well. This baseline measure is also subject to machine or technical error from the specific equipment being used. Positive controls are subject to natural biological variation from the particular cell line being used. Population luminescence after treatment (relative to the positive and negative controls) is the quantity of interest for each treatment microwell. That is, we seek to estimate the survival rate of the population after treatment: populations close to the negative control luminescence will have mostly died; populations close to the positive control luminescence will have mostly survived; in-between levels indicate some partial survival.

We aim to build a model that treats the molecular covariates as features that potentially convey predictive information about sensitivity and resistance to different therapies. For $963$ of the cell lines, we have molecular information about gene mutations, copy number variations (CNVs), and gene expression. We preprocessed the mutations and CNVs to filter down to genes that are recurrently observed as altered in large-scale observational cancer studies. This preprocessing is necessary, as adjacent CNVs are highly correlated and many nucleotide mutations occur in genes that are known to have no impact on functionality. Biological knowledge from pre-existing databases is leveraged here to filter genes down to a more informative set (see Supplementary material for details). After preprocessing, we are left with $5822$ binary gene mutations and $234$ copy number counts; we keep all $17271$ gene expression covariates. For a handful of cell lines ($109$), no molecular information is available; we treat these as missing data. Almost all drugs have been screened against all cell lines, yielding a dataset of $225385$ (cell line, drug) experiments. We treat all missing experiments and all missing molecular features as missing at random. 




%% file: related_work.tex
\subsection{Related work}
\label{subsec:introduction:related-work}
Several methods for dose-response modeling have been proposed in the recent literature. \citet{low-kam:etal:2015:bart-dose-response} proposed a Bayesian regression tree model for estimating toxicity at different levels of certain nanoparticles. \citet{wheeler:2019:bayesian-additive-dose-response} modeled dose-response with molecular descriptors via additive Bayesian splines. Both of these methods, while capable of working from raw data, do not scale to the massive size of our data, where we have tens of thousands of features.

Several machine learning approaches to dose-response prediction in cancer experiments have also been proposed~\citep[e.g.][]{menden:etal:2013:dr-nn-and-rf,ammad:etal:2016:dr-kernel-bayes-mf,rampasek:etal:2017:dr-vae}. These models seek to predict a single summary statistic (usually the area under the curve or the IC50) of a parametric model fit to the more complicated raw data. Predicting this summary statistic treats the preprocessed model as a ground truth label, making it difficult to know whether improved predictive performance is because the model has learned something about the actual response of the cancer cells, or simply has ``fit the fit'' by better matching the predictive model to the parametric model. Here we seek to work end-to-end, predicting directly from the raw data. This prevents us from fairly comparing against most past machine learning approaches that require a clearly-defined covariate and scalar response setup. The upside is that modeling the entire experiment enables us to account for measurement error, an important source of noise in scientific data~\citep{loken:gelman:2017:measurement-error} that has historically been overlooked in cancer screens. Substantial measurement error (\cref{subsec:model:batch_effects}) and outcome uncertainty (\cref{subsec:model:natural_variation}) are present in our dataset; \cref{sec:benchmarks} shows the benefits of the end-to-end approach, as well as a comparison to the state-of-the-art approach in computational biology~\citep{vis:etal:2016:sanger-dose-response}.

%% file: details.tex
\section{Bayesian dose-response model}
\label{sec:model}
\subsection{Generative model}
\label{sec:model:generative_model}
Consider $N$ cell lines and $M$ drugs. Each cell line $i = 1, \ldots, N$ is tested against each drug $j = 1, \ldots, M$, at dose levels $t = 1, \ldots, D$. The study consists of plates $\ell = 1, \ldots, L$; we denote by $\ell(i, j)$ the plate on which a specific $(i,j)$ pair was tested. The result of a plate experiment is fluorescence count data $(\mathbf{r}_{\ell}, \mathbf{q}_{\ell}, Y_{\ell})$ where $\mathbf{r}_{\ell} = (r_{\ell 1}, \ldots, r_{\ell R})$ are the $R$ negative control measurements used to estimate the baseline fluorescence bias, $\mathbf{q}_{\ell} = (q_{\ell 1}, \ldots, q_{\ell Q})$ are the $Q$ positive control measurements used to estimate the fluorescence of a population of cells when no effective treatment has been applied, and $Y_{\ell}$ are the treatment well measurements for each of the drugs on plate $\ell$. Since each $(i,j)$ pair is tested only on a single plate, we index treatments by their cell line, drug, and dose levels, respectively; thus, $\mathcal{Y} \in \R^{N \times M \times D}$ is a 3-tensor and $y_{ijt}$ represents the result of treating cell line $i$ with drug $j$ at dose level $t$. Each cell line $i$ has an associated vector of mutation, copy number variation, and expression covariates $X_i$.

We propose a generative model of dose-response,
\begin{equation}
\label{eqn:generative_model}
\begin{aligned}
r_{\ell k} &\sim& \text{Poisson}(c_\ell) \\
q_{\ell k} &\sim& \text{Poisson}(\lambda_\ell + c_\ell)\\
y_{ijt} &\sim& \text{Poisson}(\tau_{ijt}\lambda_{\ell(i,j)} + c_{\ell(i,j)}) \\
\lambda_{\ell} &\sim& \text{Gamma}(a_\ell, b_\ell) \\
\tau_{ijt} &=& \frac{1}{1+\exp(-\beta_{ijt})} \\
\boldsymbol\beta_{ij} &\sim& \text{MVN}^{+}(\boldsymbol\mu_{ij}, \Sigma_j)\\
\boldsymbol\mu_{ij} &=& f(X_i, j; \theta) 
\end{aligned}
\end{equation}
The plate-specific negative and positive control fluorescent counts, $r_{\ell k}$ and $q_{\ell k}$, are modeled as random variables for which we receive $R$ and $Q$ i.i.d. observations, respectively. The positive control fluorescence distribution is a function of the baseline fluorescence from the machine ($c_\ell$) and the growth rate ($\lambda_\ell$) of the cell line on plate $\ell$. The outcome of experiment $y_{ijt}$ is a sample from the positive control distribution, but with a treatment effect $\tau_{ijt}$ that reduces the growth rate of the cells.

The treatment effect $\tau_{ijt}$ has the direct interpretation as drug $j$ killing $(1-\tau_{ijt}) \times 100\%$ of cell line $i$ on average when applied at dose level $t$. The vector of responses $\boldsymbol\tau_{ij}$ is the dose-response curve -- that is, it represents the percentage of cells that survive at different dose levels; this is the primary quantity of interest in every experiment. Discovering effective drugs corresponds to finding curves that show sensitivity in some subset of cell lines, indicating the therapy is likely targeting some molecular property of the cells.

The dose-response curve is modeled through a latent constrained multivariate normal (MVN). The logistic transform from the MVN variable $\beta_{ijt}$ to the effect $\tau_{ijt}$ constrains effects to be in the $[0,1]$ interval, corresponding to expected cell survival percentage. The MVN is constrained to the monotone-increasing half-space, encoding that the drug effect can only become stronger as the dose increases. The mean response is a monotone function $f$ of the molecular features for the target cell line and the ID of the drug to be applied.



The model for $\boldsymbol\tau$ encodes several scientific assumptions. We assume no drug has a positive effect on any cell --- that is, no drug actually encourages growth. There are two biological motivations for this assumption. First, cancer cells are generally defined by uncontrolled proliferation. Cultivating non-cancerous cells \textit{in vitro} is extremely difficult as most cells induce apoptosis (cell suicide) outside of their host environment. There is little room left biologically for a drug to encourage the cancer cell lines to grow even more. Second, the drugs chosen for the GDSC experiment have all been selected for their ability to stress and kill cells. A large portion of the drugs are established cancer therapeutics that are designed to kill cells by targeting pathways recurrently found altered in certain cancers. In addition to assuming all drugs do not encourage growth, we also assume that toxicity only increases with drug concentration. This assumption is again based on the notion of cancer drugs being highly toxic. We check these assumptions empirically in the Supplementary material. 

\subsection{Overview of approximate inference approach}
\label{subsec:model:inference}
In practice, simultaneous estimation of all parameters in \eqref{eqn:generative_model} is 
computationally
infeasible.
Furthermore, we discovered contamination in the GDSC data that must be handled carefully (see the Supplementary material for details).
Instead, the prior parameter estimates $(\hat{a}_\ell, \hat{b}_\ell, \hat{c}_\ell, \hat{\boldsymbol\mu}_{ij}, \hat{\theta}, \widehat{\Sigma}_j)$ are obtained through a stepwise procedure. At a high level, there are four main steps:
\begin{enumerate}
    \item Baseline fluorescence bias rate $\hat{c}_{\ell}$ is estimated from negative controls. These controls contain systematic biases due to technical error; we detail a denoising approach for negative controls in \cref{subsec:model:batch_effects}.
    \item Positive control priors $\hat{a}_\ell$ and $\hat{b}_\ell$ are estimated by maximum likelihood for each plate in \cref{subsec:model:natural_variation}.
    \item The black box predictive model $f$ is chosen to be a neural network; parameters $\hat{\theta}$ are estimated in \cref{subsec:model:deep_learning} by maximizing the marginal log likelihood of the data with fixed control priors and an identity covariance matrix.
    \item Drug-specific dose-response correlation structure $\widehat{\Sigma}_j$ is estimated in \cref{subsec:model:uncertainty_estimation} through cross-validation and maximum likelihood with fixed prior means $\hat{\boldsymbol\mu}$.
\end{enumerate}
The final model enables approximate Bayesian posterior inference on $\boldsymbol\tau_{ij}$, the dose-response curve. The remainder of this section details the four estimation steps in our estimation procedure.

\subsection{Batch effects}
\label{subsec:model:batch_effects}
An assumption in high-throughput cancer drug analyses is that each individual well is independent of the other wells on the same plate, and each plate is independent of subsequent plates. There is growing concern about both assumptions among biologists. Recent studies suggest there is substantial spatial bias in microwell assays~\citep{lachmann:etal:2016:hts-spatial-bias,mazoure:etal:2017:hts-spatial-bias}, inducing dependence among individual wells on the same plate. We generally label this ``cross-contamination'' since microwell results are spilling over into neighboring wells. This cross-contamination leads us to throw out several negative control wells in each plate; see the Supplementary for details on removing cross-contaminated wells. After decontamination, some plates are left with as few as $5$ negative control wells.

\input{fig_control_estimates}

The second assumption is that each plate of observations is independent, regardless of when or where it was tested; as we show, this assumption is also violated for the GDSC data. \cref{fig:control_estimates} (left) shows one of the four (site, assay) stratifications, with negative control medians for each plate in gray.\footnote{We use medians rather than means to avoid spurious contamination not removed in the decontamination step.} The x-axis in each subplot is the date ID of the plate (the date when the plate was screened); all dates are relative to the first day of the study. There is clear visual evidence of temporal dependence, with trends followed by sharp discontinuities.

This temporal dependency is well-established in the HTS literature~\citep{johnson:etal:2007:batch-effects,leek:etal:2010:batch-effects} and falls broadly under the term ``batch effects.'' These are technical artifacts that cause otherwise-independent experiments to yield dependent results. Myriad causes can introduce temporal batch effects: using the same test site, the same equipment, preparation by the same technician, or even conducting the experiment at the same ambient temperature. Generally batch effects are seen as a nuisance that reduces the signal in experimental data and must be removed as a preprocessing step, if possible.



We instead use the temporal batch effects to our advantage when estimating the negative controls. We leverage the inter-well dependence through an approximate Bayes procedure that shrinks the differences between median estimates of negative control wells for plates screened on similar days. We use the relevant portion of the generative model for a single negative control well,
\begin{equation}
\label{eqn:hierarchical_neg_control}
\begin{aligned}
r_{\ell k} &\sim& \text{Poisson}(c_\ell) \\
c_\ell &\sim& \text{Gamma}(v_{d(\ell)}, w_{d(\ell)}) \, ,
\end{aligned}
\end{equation}
where $d(\ell)$ denotes the specific day $d$ that plate $\ell$ was screened. We use the median observation $\widetilde{c}_\ell$ as a noisy approximation to the true rate. We then fit a time-evolving density to model the prior distribution of the medians on each day,
\begin{equation}
\label{eqn:trend_filtering_density}
\begin{aligned}
& \underset{\mathbf{v}, \mathbf{w} \in \R^+}{\text{minimize}}
& & 
-\sum_d \sum_{\ell \in \{\ell \colon d = d(\ell)\}} \log (\text{Gamma}(\widetilde{c}_\ell; v_d, w_d)) + \rho_1 \vnorm{\Delta^{(1)}\frac{\mathbf{v}}{\mathbf{w}}}_1 + \rho_2 \vnorm{\Delta^{(0)}\frac{\mathbf{v}}{\mathbf{w}^2}}_1 \, ,
\end{aligned}
\end{equation}
where $\Delta^{(k)}$ is the $k^{\text{th}}$-order trend filtering matrix~\citep{tibshirani:2014:trend-filtering} using the falling factorial basis~\citep{wang:etal:2014:falling-factorial-basis} to handle the irregular grid of days. The solution to \eqref{eqn:trend_filtering_density} finds densities that are piecewise-linear in their mean and piecewise constant in their variance. 

The regularization parameters $\rho_1$ and $\rho_2$ are chosen via 5-fold cross-validation and the model is fit with stochastic gradient descent. Since the counts are large, we use a normal approximation to the gamma, parameterized in natural parameter space for computational convenience; shape and rate parameters are reconstructed from the mean and variance of the normal. Since \eqref{eqn:trend_filtering_density} is non-convex, we only find a local optimum, but empirically we observe good fits across a wide array of simulated data. The orange line and bands in the left panel of \cref{fig:control_estimates} show the results for one stratification, where the outliers have been shrunk substantially.

Given the learned prior $(\hat{v}_d, \hat{w}_d)$ from \eqref{eqn:trend_filtering_density}, we calculate the \textit{maximum a posteriori} (MAP) estimate for the negative control mean,
\begin{equation}
\label{eqn:map_neg_control}
\hat{c}_\ell = \frac{\hat{v}_{d(\ell)} + \sum_k r_{\ell k}}{\hat{w}_{d(\ell)} + \sum_k 1} \, .
\end{equation}
While there is technical variation in the bias rate, it is relatively small after corrections and thus as a practical matter we simply use the MAP estimate for the Poisson rate of the negative controls. Using the MAP estimate with a fixed $\hat{c}_\ell$ simplifies downstream inference by removing the need to numerically integrate out the negative control.

\subsection{Natural variation in cell line growth}
\label{subsec:model:natural_variation}
Even under ideal conditions without any batch effects or spatial plate bias, cell population growth and response exhibits a large degree of natural variation. For instance, \cref{fig:control_estimates} (right) shows the distribution of the positive control wells for one example plate. The variance in this cell line is so large that a decrease in fluorescence of even $50\%$ compared to the mean would not be highly unlikely. Furthermore, population growth variance varies substantially between cell lines, testing sites, and assay types.


Since we have a reasonably large number of positive control wells on each plate, we estimate the population directly,
\begin{equation}
\label{eqn:map_pos_control}
\begin{aligned}
\underset{a, b \in \R^+}{\text{maximize}} & \prod_q \int \text{Poisson}(q; \lambda + \hat{c}_\ell) \text{Gamma}(\lambda; a, b) d\lambda \, .
\end{aligned}
\end{equation}
The integral in \eqref{eqn:map_pos_control} can be resolved analytically to be an incomplete gamma; however, we found a finite grid approximation to be more numerically stable. The maximum likelihood problem is also nonconvex, so we again rely on a local optima approximation found via a sequential least squares solver. We found the fits in simulation to be close to the ground truth when using the same number of control replicates as in the GDSC data. The orange line in the right panel of \cref{fig:control_estimates} shows the resulting fit from optimizing \eqref{eqn:map_pos_control} on the example plate.

\subsection{Fitting the black box response prior}
\label{subsec:model:deep_learning}
We use a deep neural network for $f$, parameterized by weights $\theta$. The molecular feature $X_i$ is passed through a $1000\times200\times200\times(265\times9)$ feed forward network.
The $9$ outputs $\boldsymbol\phi_{ij}$ for each of the $265$ drugs are then constrained to be monotone,
\begin{equation}
\label{eqn:monotone_nn_output}
\mu_{ijt} = \phi_{ij9} + \sum_{9 > k \geq t} \log(1+\exp(\phi_{ijk})) \, ,
\end{equation}
where the right-hand side of \eqref{eqn:monotone_nn_output} is the cumulative sum of the softplus operator applied to the raw outputs from dose $t$ upward, allowing the maximum dose level to set the offset. We fix $\widehat{\Sigma}_j$ to the identity matrix; we found the results did not improve by using off-diagonal terms.

We optimize $\theta$ by maximizing the log-likelihood of the data,
\begin{equation} 
\label{eqn:objective_function}
\begin{aligned}
\mathcal{L}(\theta) &= \sum_{i=1}^N \sum_{j=1}^M \log p \left(y_{ij} \mid f(x_{ij}; \theta), \widehat{\Sigma}, \hat{a}, \hat{b}, \hat{c} \right) \\
&= \sum_{i=1}^N \sum_{j=1}^M \log \int \left[ \prod_t p \left(y_{ijt} \mid \beta_{ijt}, \hat{a}, \hat{b}, \hat{c} \right)\right]
p(\beta_{ij} \mid f(x_{ij}, \theta), \widehat{\Sigma}) \, \mathrm{d} \beta_{ij} \, ,
\end{aligned}
\end{equation}
where
\begin{align*}
p \left(y_{ijt} \mid \beta_{ijt}, \hat{a}, \hat{b}, \hat{c} \right)  &=
\int \mathrm{Po}(y_{ijt} \mid \sigma(\beta_{ijt}) \lambda_{ijt} + \hat{c}) \, \mathrm{Ga}(\lambda_{ijt} \mid \hat{a}, \hat{b}) \, \mathrm{d} \lambda_{ijt} \, .
\end{align*}
We approximate the inner integral in \eqref{eqn:objective_function} with a numeric grid over the values of $\lambda_{ijt}$.

We split the data into $10$ cross-validation (CV) folds. For each fold, we use $90\%$ of the in-sample data as training and $10\%$ as validation for early-stopping. We optimize \eqref{eqn:objective_function} using RMSprop~\citep{tieleman:hinton:2012:rmsprop} for $30$ epochs with $10\times265$ samples per mini-batch. We check empirical risk on the validation set after every epoch and keep the best model over the entire run. The final model predictions on the out-sample (test) fold are then used for evaluation of the prior in \cref{sec:benchmarks}.

\subsection{Estimating dose-response covariance and posterior dose-response curves}
\label{subsec:model:uncertainty_estimation}
After fitting $\hat{\theta}$, we reuse the validation set to estimate the covariance matrix for each drug. Fixing the predicted means for each test set, we estimate the drug dose-response covariance as the maximum likelihood estimator,
\begin{equation}
\label{eqn:covariance_mle}
\begin{aligned}
\hat{\boldsymbol\tau}_{ij} &=& \textrm{max}\left(0, \textrm{min}\left(1, \frac{(\mathbf{y}_{ij} - \hat{c}_{\ell(i,j)})}{\hat{a}_{\ell(i,j)} \times \hat{b}_{\ell{(i,j)}}}\right)\right)\\
\hat{\boldsymbol\beta}_{ij} &=& \log\left(\frac{\hat{\boldsymbol\tau}_{ij}}{1-\hat{\boldsymbol\tau}_{ij}}\right) \\
\widehat{\Sigma}_j &=& (\hat{\boldsymbol\beta}_{\cdot j} - \hat{\boldsymbol\mu}_{\cdot j})^\intercal(\hat{\boldsymbol\beta}_{\cdot j} - \hat{\boldsymbol\mu}_{\cdot j}) \, ,
\end{aligned}
\end{equation}
where $\hat{\boldsymbol\tau}$ is the empirical estimate of cell survival (i.e. without prior information from the features or any constraint on the curve shape).

The prior estimates are then fixed and the posterior distribution of $\boldsymbol\tau$, the latent dose-response rate for each experiment, is estimated via Markov chain Monte Carlo (MCMC).
We evaluated two MCMC methods:
1) a fully-conjugate Gibbs sampler implemented via Polya-Gamma augmentation~\citep{polson:etal:2013:polya-gamma} where the sampled posterior MVN logits are projected to be monotone using the pool adjacent violators algorithm as in~\citep{lin:dunson:2014:monotone-gp}, and 2) rejection sampling with an elliptical slice sampling~\citep{murray:etal:2010:elliptical-slice-sampling} proposal. We found the Gibbs sampler to have high sample complexity due to the need to sample both $\lambda_{ijt}$ and $\tau_{ijt}$, where the value of one tightly constrains the distribution over the other. The elliptical slice sampler, by contrast, approximated the posterior better with fewer samples and only required a single MCMC chain. We ran the elliptical slice sampler for $2000$ iterations with the first $500$ iterations discarded as burn-in samples.

%% file: fig_control_estimates.tex
\begin{figure}[ht]
\centering
\subfigure[Negative control uncertainty]{\includegraphics[width=0.5115\textwidth]{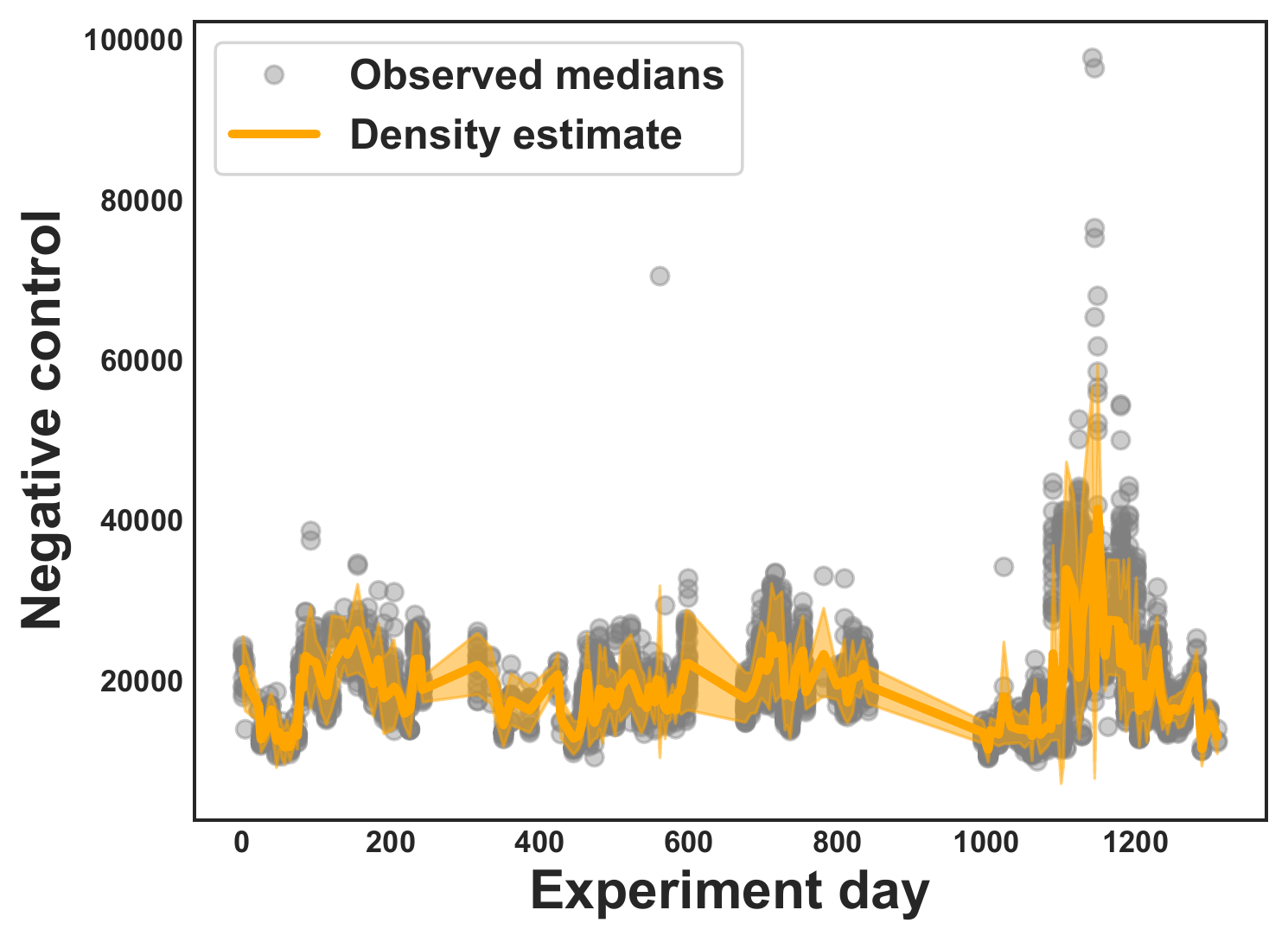}}
\subfigure[Positive control uncertainty]{\includegraphics[width=0.4785\textwidth]{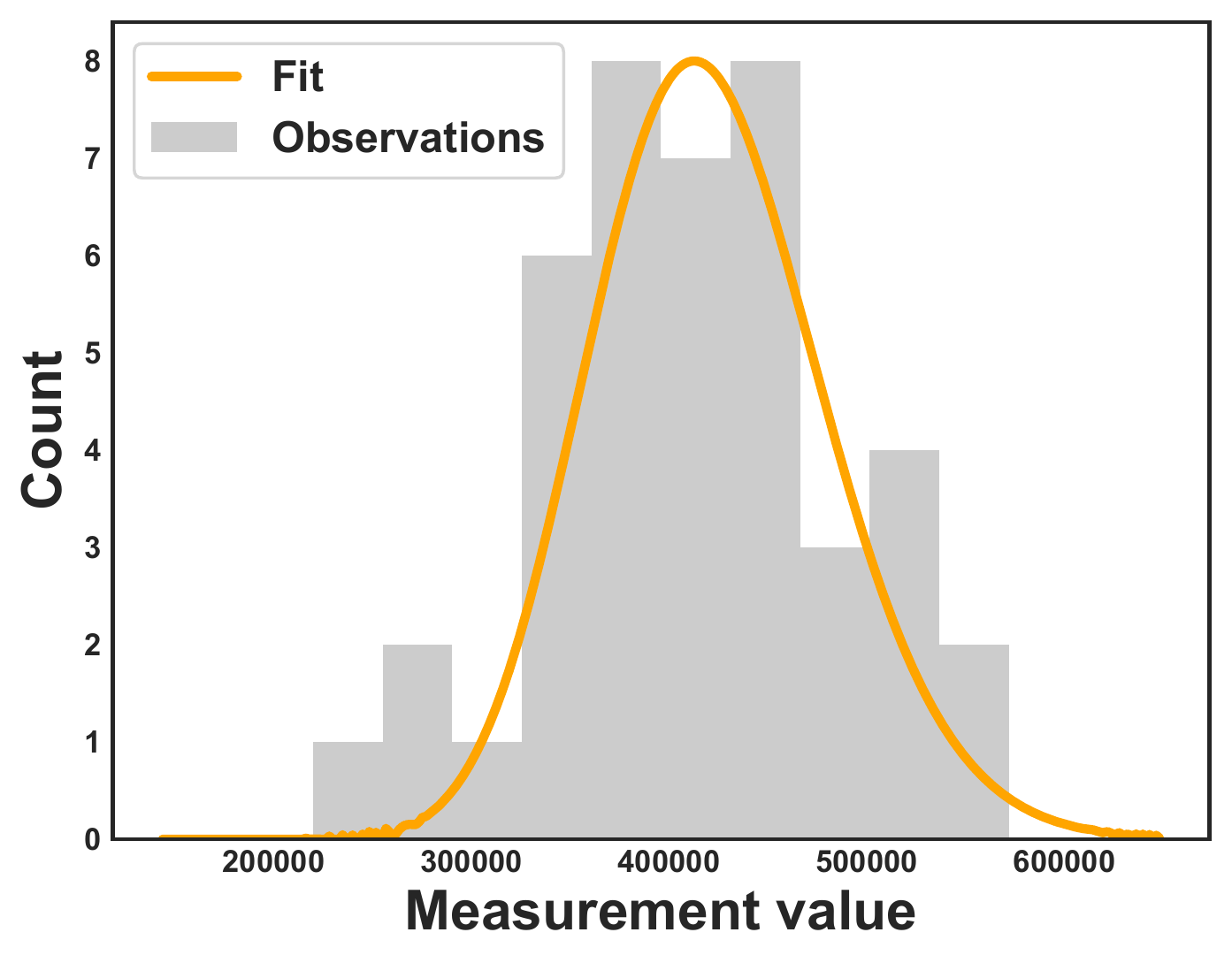}}
\caption{\label{fig:control_estimates} Left panel: Time-evolving estimate of the negative control density for an example testing site and assay type. The x-axis is the date a plate was screened, relative to the start of the study; y-axis is the median control well measurements after removing contaminated wells. The trend filtering density regression fit is in orange (line: mean, bands: $90\%$ regions). Right panel: Maximum likelihood estimate of an example positive control density, after fitting the negative control MAP estimate.}
\end{figure}

%% file: benchmarks.tex
\section{Model comparison and evaluation}
\label{sec:benchmarks}
\subsection{Baseline approach}
\label{subsec:benchmarks:baseline}
We compare the proposed Bayesian approach to the existing dose-response curve fitting pipeline~\citep{vis:etal:2016:sanger-dose-response} used for the GDSC dataset. The pipeline is similar to those used in other high-throughput cancer drug screening analyses~\citep[e.g.,][]{barretina:etal:2012:ccle}. Unlike in the Bayesian approach, the baseline method does not account for measurement error, model natural variation, or quantify its uncertainty.

First, negative and positive controls are averaged to obtain $\bar{r}_\ell$ and $\bar{q}_\ell$. These are point estimates of baseline fluorescence when all cells die or survive, respectively, on plate $\ell$. The control point estimates are then used to calculate the expected percentage of cells that survived each treatment experiment,
\begin{equation}
\label{eqn:standard_y_estimate}
\tilde{\tau}_{ijt} = \text{max}\left(0, \text{ min}\left(1, \frac{\bar{q}_{\ell(i,j)} - y_{ijt}}{\bar{q}_{\ell(i,j)} - \bar{r}_{\ell(i,j)}}\right)\right) \, .
\end{equation}
The $\tilde{\tau}$ estimates are then treated as observations of the percentage of cells surviving. A logistic curve is fit for every $(i,j)$ pair using a multilevel mixed effects model,
\begin{equation}
\label{eqn:sanger_model}
\begin{aligned}
\hat{\tau}_{ijt} &=& \frac{1}{1+e^{-\frac{t - \zeta_1 + \xi_{1i} + \xi_{1ij}}{\zeta_2 + \xi_{2i}}}} \\
[\xi_{1i}, \xi_{2i}] &\sim& \mathcal{N}(0, \Psi) \\
\xi_{1ij} &\sim& \mathcal{N}(0, \nu^2) \, ,
\end{aligned}
\end{equation}
where $\xi_{1i}$ and $\xi_{2i}$ enable the model to share statistical strength between drugs tested on the same cell line; the model is fit by maximum likelihood estimation. Curves are summarized by integrating out the dose parameter $t$ to obtain a summary statistic, such as the estimated concentration required to kill $50\%$ of cells (IC50). Log-IC50 values are used as targets for predictive modeling of molecular features. For each drug, an elastic net~\citep{zou:hastie:2005:elastic-net} model is fit with hyperparameters chosen through cross-validation.

\subsection{Performance comparison}
\label{subsec:benchmarks:quantitative}
We compare the Bayesian model in \cref{eqn:generative_model} to the above baseline approach. To compare the two methods, we consider the task of imputing a missing dose given observations of the other dose levels in the experiment. This checks the ability of the model to capture the shape of the dose-response curve. For each (cell line, drug) experiment, we hold out one dose level at random and treat it as missing data that must be imputed. Error is measured in terms of variance-adjusted raw count values on the held out data,
\begin{equation}
\label{eqn:mse_criteria}
\textrm{err}_{ijt} = \left(\frac{y_{ijt} - \hat{y}_{ijt}}{\textrm{stdev}(\mathbf{q}_{\ell(i,j)})}\right)^2 \, ,
\end{equation}
where $\hat{y}_{ijt}$ is the model prediction and $\textrm{stdev}(\mathbf{q}_{\ell(i,j)})$ is the standard deviation of the raw positive controls. For the baseline, raw predictions are backed out from \eqref{eqn:standard_y_estimate} and \eqref{eqn:sanger_model},
\begin{equation}
\label{eqn:baseline_yhat}
\hat{y}^{\text{(baseline)}}_{ijt} = \hat{\tau}^{\text{(baseline)}}_{ijt}(\bar{q}_{\ell(i,j)} - \bar{r}_{\ell(i,j)}) + \bar{r}_{\ell(i,j)} \, .
\end{equation}
For the Bayesian model, we use a MAP estimate of the raw count,
\begin{equation}
\label{eqn:bayesian_yhat}
\hat{y}^{\text{(Bayes)}}_{ijt} = \hat{\tau}^{\text{(Bayes)}}_{ijt}\times\hat{a}_{\ell(i,j)}\times\hat{b}_{\ell(i,j)} + \hat{c}_{\ell(i,j)} \, ,
\end{equation}
where $\hat{\tau}^{\text{(Bayes)}}_{ijt}$ is the posterior mean estimate of $\tau_{ijt}$ in \eqref{eqn:generative_model}. We also consider a hybrid version of the baseline that uses only the control correction technique. For this method, we replace the control means with the approximate Bayes estimates,
\begin{equation}
\label{eqn:hybrid_yhat}
\hat{y}^{\text{(hybrid)}}_{ijt} = \hat{\tau}^{\text{(baseline)}}_{ijt}\times\hat{a}_{\ell(i,j)}\times\hat{b}_{\ell(i,j)} + \hat{c}_{\ell(i,j)} \, .
\end{equation}
In the hybrid model, the controls are used somewhat differently in two ways. First, the raw controls are used by the baseline method to estimate $\hat{\tau}^{\text{(baseline)}}$ then the corrected controls are used to estimate $\hat{y}^{\text{(hybrid)}}$.


\input{tab_dose_imputation}

\cref{tab:dose_imputation} presents the results for this benchmark. Using the corrected controls in the hybrid model improves the predictions of the baseline model slightly, suggesting the correction procedures from \cref{subsec:model:batch_effects,subsec:model:natural_variation} are useful independent of the Bayesian model. However, the Bayesian model outperforms the predictions from the corrected baseline model by $\approx20\%$. We also investigated a featureless version of the Bayesian model and found similar improvements over the baseline, suggesting the improvements are due to the flexibility of the constrained multivariate normal prior. The log-linear constraint of the baseline model imposes a strong assumption that the true dose response curve has a sigmoidal shape. By contrast, the Bayesian model only assumes monotonicity and learns the shape priors from the data.

\subsection{Assessing feature importance}
\label{subsec:benchmarks:feature_ablation}
A current debate in precision oncology is whether all molecular feature types contain predictive power. Gene panels used in hospitals, such as MSK-IMPACT~\citep{cheng:etal:2015:msk-impact}, only consider mutations and copy number, while others argue that gene expression is sufficient to predict response~\citep{piovan:etal:2013:akt-inhibition,rodriguez:etal:2015:jak2-inhibition}. We investigate this here by fitting separate models for every possible subset of features (mutations, copy number variations, and expression). If a feature set does not contain worthwhile information, it will effectively introduce noise into the model and either not improve performance or lower it (e.g. due to finite samples and a nonconvex optimization procedure).

To measure predictive power of a model, we consider the task of predicting entire out-of-sample experiments at all dose levels. We again measure error on variance-adjusted raw count predictions. However, the curve prediction task is a prior predictive check of the model, rather than a posterior one. This is a more challenging task, as the model does not see any outcomes from the specific experiment when making predictions. We use the logistic-transformed mean as the predicted drug effect,
\begin{equation}
\label{eqn:prior_yhat}
\hat{y}^{\text{(prior)}}_{ijt} = \frac{\hat{a}_{\ell(i,j)}\times\hat{b}_{\ell(i,j)}}{1+e^{-f^{+}_t(X_i, j; \hat{\theta})}} + \hat{c}_{\ell(i,j)} \, .
\end{equation}
The approach of~\citet{vis:etal:2016:sanger-dose-response} does not provide a way to make raw predictions from feature subsets; we therefore only consider the Bayesian model for this task.

\input{tab_curve_prediction}

We evaluate the predicted priors from the held out cross-validation folds, across all folds. We measure mean error by first taking the average error on the entire curve, then averaging across all curves. \cref{tab:curve_prediction} shows the curve prediction results. The model generally has lower error as more feature subsets are included, suggesting that all three feature subsets add predictive value.

\subsection{Qualitative evaluation}
\label{subsec:benchmarks:qualitative}
In addition to the quantitative results above, we also evaluate whether the model recapitulates known biology in its learned prior. This qualitative check provides reassurance 
that the patterns discovered by the model are reliable and not likely just functions of undetected experimental artifacts. We generated marginal prior predictive curves with different subsets of drugs and cell lines. The subsets of cells are chosen based on features that are known to be targeted by certain drugs. Drugs that target one subset should produce more sensitive marginal predicted response curves.

\input{fig_known_biology}

\paragraph{BRAF inhibitors:} $6$ of the drugs in the dataset are labeled as \textit{BRAF inhibitors}. These drugs all target a well-studied, commonly-mutated oncogene called BRAF. Mutations in this gene, particularly mutations in the codon 600 (V600E), lead to activation of the MAPK (MEK/ERK) signaling pathway in tumor cells. As shown in the top left panel of \cref{fig:known_biology}, the model successfully captures the differential efficacy of BRAF inhibitors between cell lines possessing mutant and wild type variants.

\paragraph{MAPK signaling:} $18$ of the drugs in the dataset target signaling in the \textit{MAPK} pathway. The MAPK signaling pathway is involved in cell growth and proliferation and one of the most commonly activated pathway across tumors. KRAS and NRAS are members of the RAS gene family, which constitute central nodes in the MAPK pathway. RAS proteins activate other proteins that in turn drive cell growth. Mutations in KRAS and NRAS can lead them to be permanently activated, causing continual growth signaling downstream in the MAPK pathway. The drugs we consider here target the MAPK pathway in various ways to downregulate its activity; specifically, the drugs included here target BRAF, RSK2, ERK1/2/5, and MEK1/2/5. The top right panel of \cref{fig:known_biology} shows that the model successfully predicts more sensitivity to these drugs when a cell line has a mutation in one of these two RAS genes.

\paragraph{Nutlin-3(a):} The drug Nutlin-3(a) binds to and suppresses MDM2. MDM2 is a negative regulator of P53. TP53, the gene that codes for the P53 protein, is the most commonly mutated gene in cancer and is known as the ``guardian of the genome'' for the multiple roles that plays regulating the damage to DNA. When DNA damage is found, TP53 will initiate a transcriptional program to attempt to repair the damage or, if the damage is too severe, will initiate apoptosis and force the cell to die. When TP53 is mutated or suppressed, these safety mechanisms are disabled and they allow downstream damage to occur unchecked, leading to oncogenic growth. MDM2 targets TP53 for proteosomal degradation and when elevated can suppress TP53 entirely. Thus, in order for Nutlin-3(a) to be effective, we need two conditions to be true: 1) MDM2 needs to be elevated above normal levels such that it functionally inactivates TP53, and 2) TP53 must not be mutated, such that if MDM2 is suppressed then TP53 will be functional and capable of initiating apoptosis. In this case, we consider a cell line to have an elevated level of MDM2 if it is expressed at least one standard deviation above the mean expression level in the dataset. The bottom left panel of \cref{fig:known_biology} confirms that the model predicts this nonlinear efficacy pattern.

\paragraph{PI3K inhibitors:} There are $24$ drugs in the dataset that are \textit{PI3K inhibitors}. These drugs broadly target the PI3K-AKT-mTOR pathway, which plays a prominent role in cell growth and division. Cells that exhibit oncogenic malfunctions in this pathway are characterized by two common hallmarks: 1) a loss or mutation of the tumor suppressor gene PTEN, and 2) elevated levels of the oncogene PIK3CA. As in the Nutlin-3(a) scenario, these two show a nonlinear interaction that is replicated by the learned prior, as seen in the bottom right panel of \cref{fig:known_biology}.\\

The qualitative checks above provide a useful reassurance, but they are not foolproof. Namely, these are checks about marginal feature importance. Most features are highly correlated with each other. This correlation may lead a marginal test to indicate the model is capturing biological knowledge about an important feature when it is really learning information about other features that are correlated with the chosen examples. As a final validation of our approach, we use the posterior dose-response curves to test for \textit{conditional} feature importance.

%% file: tab_dose_imputation.tex

\begin{table}[th]
\centering
\begin{small}\begin{tabular}{|p{16mm}|c|c|c|c|c|c|c|c|c|c|}
\hline
& \textbf{(min)} & \multicolumn{7}{|c|}{\textbf{Dose level}} & \textbf{(max)} & \\ \cline{2-10}
\textbf{Model} & \textbf{0} & \textbf{1} & \textbf{2} & \textbf{3} & \textbf{4} & \textbf{5} & \textbf{6} & \textbf{7} & \textbf{8} & \textbf{All} \\ \hline
Baseline       & 1.00 & 1.00 & 1.00 & 1.00 & 1.00 & 1.00 & 1.00 & 1.00 & 1.00 & 1.0 \\
Hybrid         & 1.04 & 1.00 & 1.04 & 0.97 & 0.95 & 0.91 & 0.90 & 0.93 & 1.03 & 0.98 \\ \hline
Bayesian       & 0.92 & 0.86 & 0.71 & 0.69 & 0.67 & 0.79 & 0.77 & 0.96 & 0.86 & 0.82 \\ \hline
\end{tabular}
\end{small}
\caption{\label{tab:dose_imputation} Mean squared error results on the single-dose imputation benchmark relative to the basline model from \citet{vis:etal:2016:sanger-dose-response}. The baseline is slightly improved by using corrected controls (Hybrid), but overall is not flexible enough to fully model the observed dose-response curves; the Bayesian model outperforms both pipelined approaches.}
\end{table}

%% file: tab_curve_prediction.tex

\begin{table}[th]
\centering
\begin{small}\begin{tabular}{|
p{15mm}|c|c|c|c|c|c|c|c|c|c|}
\hline
& \textbf{(min)} & \multicolumn{7}{|c|}{\textbf{Dose level}} & \textbf{(max)} & \\ \cline{2-10}
\textbf{Model} & \textbf{0} & \textbf{1} & \textbf{2} & \textbf{3} & \textbf{4} & \textbf{5} & \textbf{6} & \textbf{7} & \textbf{8} & \textbf{All} \\ \hline
Mutations  & 1.06 & 1.08 & 1.07 & 1.06 & 1.04 & 1.03 & 1.02 & 1.02 & 1.00 & 1.03 \\
CNV        & 1.04 & 1.06 & 1.07 & 1.08 & 1.09 & 1.11 & 1.11 & 1.13 & 1.12 & 1.10 \\
Expression & 1.01 & 1.03 & 1.02 & 1.03 & 1.02 & 1.03 & 1.01 & 1.00 & 1.02 & 1.02 \\
Mut+CNV    & 1.05 & 1.04 & 1.06 & 1.05 & 1.04 & 1.03 & 1.01 & 1.02 & 1.00 & 1.03 \\
Mut+Exp    & 1.00 & 1.01 & 1.01 & 1.01 & 1.01 & 1.00 & 1.00 & 0.99 & 0.99 & 1.00 \\
CNV+Exp    & 0.99 & 1.00 & 1.00 & 1.01 & 1.01 & 1.01 & 1.01 & 1.00 & 1.00 & 1.01 \\ \hline
All        & 1.00 & 1.00 & 1.00 & 1.00 & 1.00 & 1.00 & 1.00 & 1.00 & 1.00 & 1.00 \\\hline
\end{tabular}
\end{small}
\caption{\label{tab:curve_prediction} Mean error results on the curve prediction benchmark relative to using all molecular features. Overall performance generally increases as more features are added, suggesting each subset conveys valuable predictive information not captured by the other two.}
\end{table}

%% file: fig_known_biology.tex
\begin{figure}[ht]
\centering
\subfigure{\includegraphics[width=0.47\textwidth]{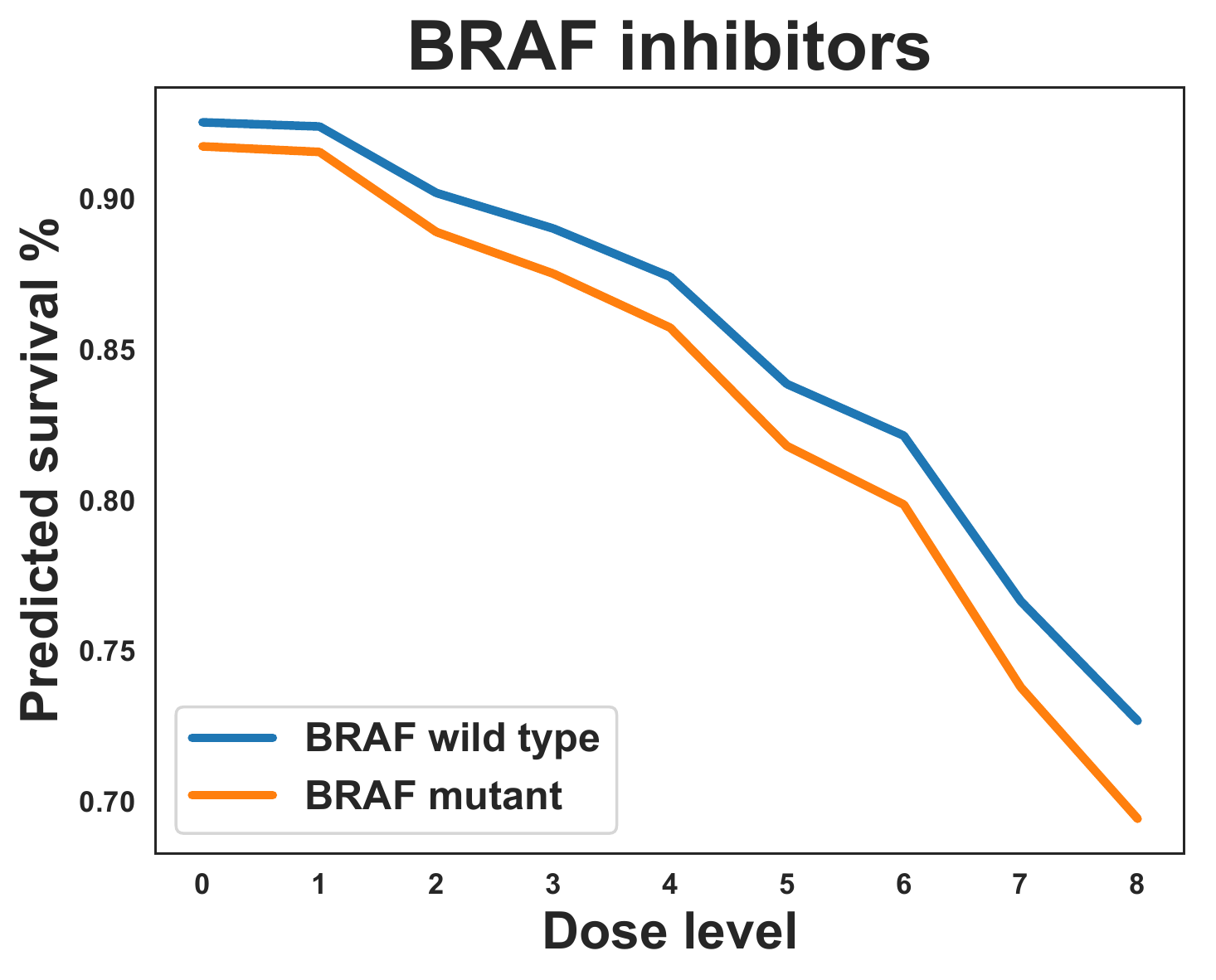}}
\subfigure{\includegraphics[width=0.47\textwidth]{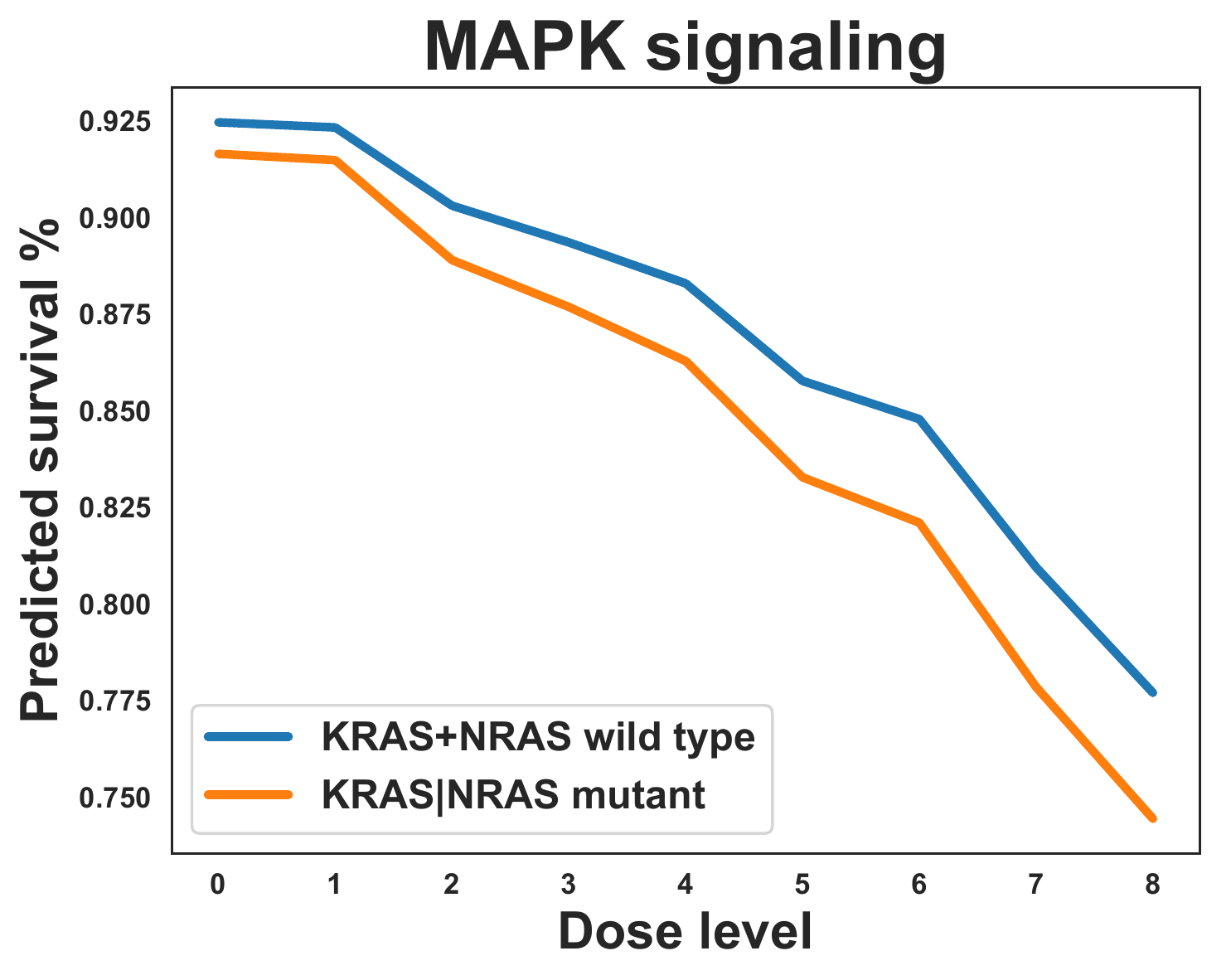}} \\
\subfigure{\includegraphics[width=0.47\textwidth]{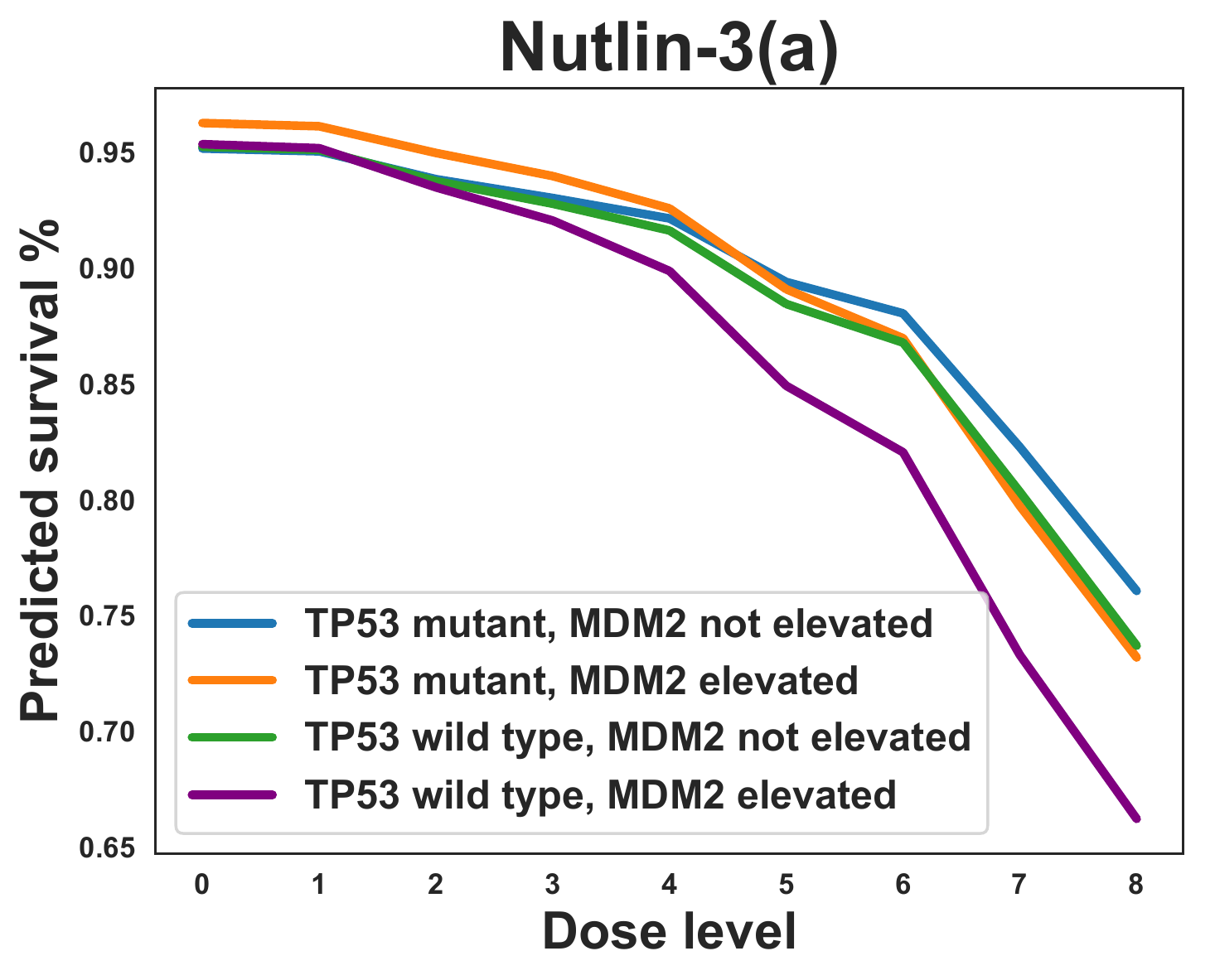}}
\subfigure{\includegraphics[width=0.47\textwidth]{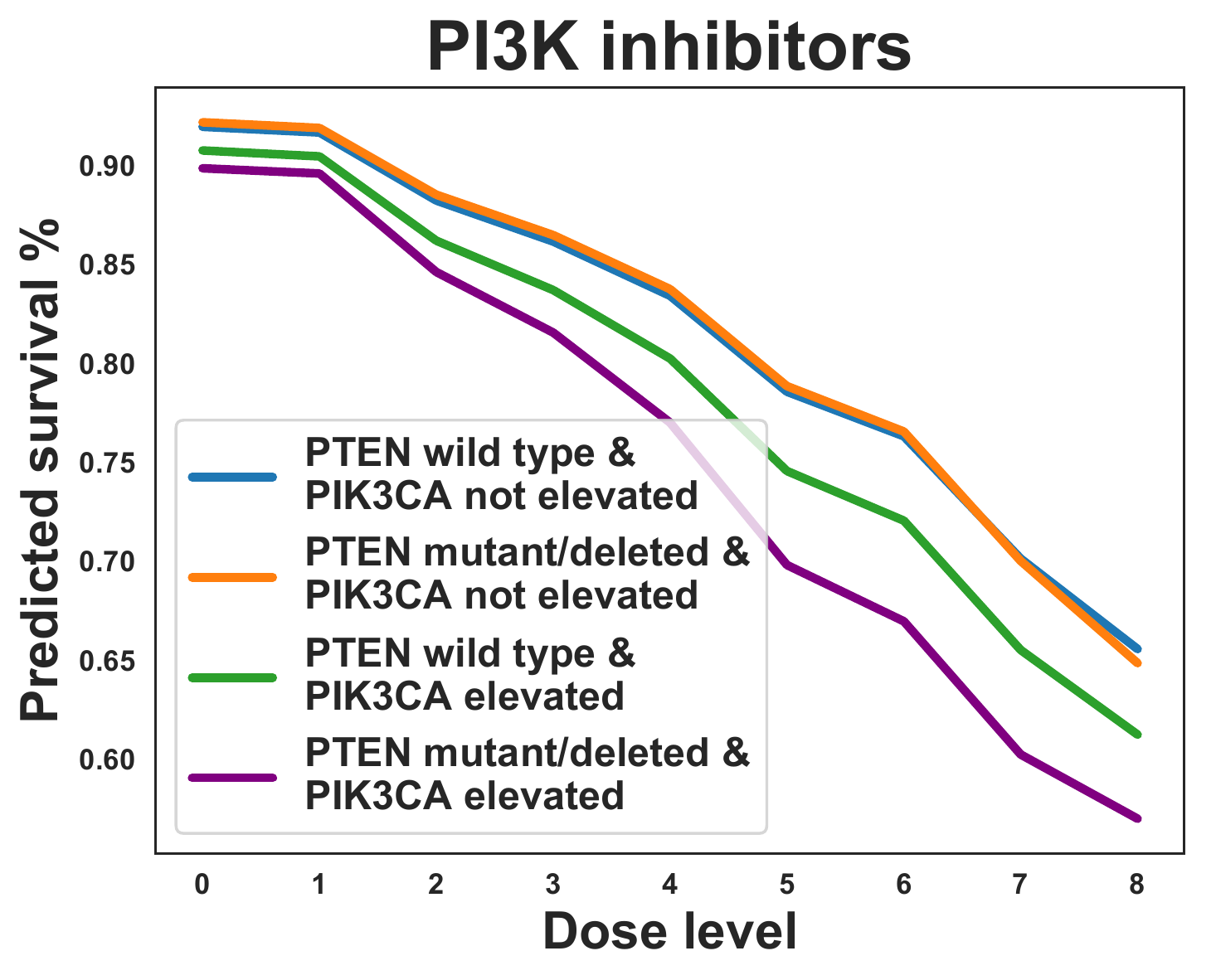}}
\caption{\label{fig:known_biology} Prior predictive checks recapitulating known biology. Top left: cell lines with and without a mutated BRAF gene, when treated with any of the six drugs designed to target such cells. Top right: Drugs targeting MAPK signaling that shut down the aberrant signaling caused by mutations in RAS-type genes. Bottom left: Nutlin-3(a) (an MDM2 inhibitor) is effective only when both MDM2 levels are high and TP53 is not mutated. Bottom right: PI3K inhibitors target cells exhibiting PTEN loss and PIK3CA elevation.}
\end{figure}

%% file: biomarkers.tex
\section{Biomarker discovery}
\label{sec:biomarkers}
\subsection{Formulating biomarker detection as a multiple testing problem}
\label{subsec:biomarkers:setup}
In addition to predictive modeling, a second goal in cancer drug screening is to detect biological markers,
e.g., genomic or transcriptomic variations in a sample,
 that allow us to predict that a 
sample will be sensitive or resistant to therapy. These biomarkers represent potential causal links that, if validated in wet lab experiments and clinical trials, can be used to craft personalize treatments for patients. Statistically, we formulate biomarker identification as a multiple hypothesis testing problem over a set of binary features. The null hypothesis is that a feature $X_g$ is conditionally independent of response to drug $Y_j$, given all other features,
\begin{equation}
\label{eqn:conditional_indep}
H_0 \colon X_g \bigCI Y_j \mid X_{-g} \, .
\end{equation}
We argue that testing for conditional independence is more appropriate for biomarker discovery than the current approaches often employed in biology. Existing approaches in the literature either test for marginal dependence~\citep{yang:etal:2012:gdsc} or use heuristic feature selection techniques like lasso regression~\citep[c.f.,]{barretina:etal:2012:ccle}. The marginal dependence case leads to many spurious false positives due to the high degree of dependence between features. Heuristic methods, on the other hand, do not offer quantification of uncertainty nor provide a way to control false positives. By contrast, conditional independence testing corrects for feature dependence and casting the problem as hypothesis testing enables us to apply existing techniques for multiple hypothesis correction to achieve finite-sample control over the target error rate (e.g. Type I error or false discovery rate).

Biomarkers are binary features. A sample is either positive or negative for a biomarker. Therefore, we convert all molecular features to binary biomarkers and test each for conditional independence. Specifically, we consider the following candidate features as potential biomarkers:
\begin{itemize}
    \item \textbf{Mutations}. These features are already binary in nature.
    \item \textbf{Copy number gain and loss}. We define copy number gain (loss) binary features as a copy number greater (less) than the median copy number for the cell line. For normal diploid human cells this is $2$, however cell lines exhibit high degrees of anuploidy; the median copy number of cell lines in the GDSC is $3$.
    \item \textbf{Over and under expression}. We define a gene to be over (under) expressed if it has expression of greater (less) than $1$ standard deviation above (below) the mean.
\end{itemize}
We discard any feature with less than $5$ positive and negative samples to 
lower the likelihood of
spurious results. The final dataset has $37176$ binary features which we test across all $265$ drugs for a total of $9.8$M hypotheses.

\subsection{Amortized conditional randomization testing}
\label{subsec:biomarkers:crts}
To test for conditional independence, we develop a new type of conditional randomization test (CRT)~\citep{candes:etal:2018:panning}. As with any randomization test, the CRT repeatedly samples from the null distribution to build up an empirical distribution of an arbitrary test statistic. The observed data can then be compared against the null samples to calculate a $p$-value for each null hypothesis.

In the conditional independence case, the null distribution is $P(X_g \mid X_{-g})$. In practice, the null distribution must be estimated from the data. With thousands of features, fitting this many models can be computationally prohibitive. Rather than fitting conditional distributions for each feature, we posit a factor model of the dataset,
\begin{equation}
\label{eqn:factor_model}
P(X) = P(Z) \prod P(X_{g} \mid Z) \, .
\end{equation}
Given the factor model assumption, we test the surrogate null hypothesis that conditions only on the latent factor,
\begin{equation}
\label{eqn:factor_indep}
\begin{aligned}
H_0 \colon X_g \bigCI Y_j \mid X_{-g} &&\preceq&& H_0 \colon X_g \bigCI Y_j \mid Z \, .
\end{aligned}
\end{equation}
If the factor model assumption holds, then the right hand side null hypothesis is a superset of the conditional independence null hypothesis. This is because $Z$ contains at least all of the information of $X_{-g}$ in \cref{eqn:factor_model}.

The factor model assumption enables estimation of the conditional model to be 
amortized: rather 
than fit individual conditional models, we can fit a single factor model to the dataset and sample from $P(X_g \mid \hat{Z})$, where $\hat{Z}$ is the estimated latent factors. Concretely, we fit a logistic factor model to the binary dataset. We include tissue type as $55$ binary columns to ensure tissue type does not act as a latent confounder. We use $k=50$ latent factors and optimize the model until convergence to a local optimum.

We calculate $p$-values by sampling up to $10^6$ null samples, calculating their test statistics, and comparing with the true test statistic. We target a $10\%$ false discovery rate using BH correction for multiple tests. To speed up computation, we early-stop $p$-values if they are larger than $10^{-h}$ for $h=1, \ldots, 5$; these larger $p$-values will not pass a BH correction step for a $10\%$ FDR, making this a conservative step.

\subsection{Safety, toxicity, and targeted efficacy}
\label{subsec:biomarkers:targeted_efficacy}
Conditional randomization testing requires the specification of a test statistic. We propose a new dose-response summary statistic: targeted efficacy. Drugs with a high targeted efficacy score for a given biomarker have a strong differential response probability when segregating the samples based on the biomarker. The statistic reports the degree to which a drug targets a subpopulation of samples with or without the biomarker, and remains innocuous for the other subpopulation.

We first define what it means for a drug to be safe or toxic. These are subjective decisions that can be informed by biological knowledge in collaboration with experts in a specific therapy or cancer. After conversations with pathologists at Columbia University Medical Center, we chose to define a drug as safe if at least $80\%$ of cells were expected to survive treatment; toxicity was chosen to be a survival rate of at most $50\%$. To calculate the safety statistic for drug $j$ at dose level $t$, we aggregate over all samples for which the target biomarker is negative,
\begin{equation}
\label{eqn:safety}
\mathrm{Safety}(\mathbf{x}_g, j, t) = \frac{1}{\vert \{ i \colon x_{ig} = 0 \} \vert } \sum_{\{ i \colon x_{ig} = 0 \}} P(\tau_{ijt} \geq 0.8 \mid \mathbf{y}_{ij}, \hat{\boldsymbol\mu}_{ij}, \widehat{\Sigma}_j, \hat{a}, \hat{b}, \hat{c}) \, .
\end{equation}
Similarly, we define toxicity as having a survival rate of at most $50\%$,
\begin{equation}
\label{eqn:toxicity}
\mathrm{Toxicity}(\mathbf{x}_g, j, t) = \frac{1}{\vert \{ i \colon x_{ig} = 1 \} \vert } \sum_{\{ i \colon x_{ig} = 1 \}} P(\tau_{ijt} \leq 0.5 \mid \mathbf{y}_{ij}, \hat{\boldsymbol\mu}_{ij}, \widehat{\Sigma}_j, \hat{a}, \hat{b}, \hat{c}) \, .
\end{equation}
Safety and toxicity probabilities are then combined to assess overall efficacy of the drug. We calculate efficacy by taking the minimum of the safety and toxicity probabilities at each dose, then reporting the score at the best observed dose,
\begin{equation}
\label{eqn:efficacy}
\mathrm{Efficacy}(\mathbf{x}_g, j) = \underset{t}{\textrm{max}}\mbox{ }\textrm{min}\left(\mathrm{safety}(\mathbf{x}_g, j, t), \mathrm{toxicity}(\mathbf{x}_g, j, t)\right) \, .
\end{equation}
Finally, a biomarker may be an indicator of sensitivity or resistance. The targeted efficacy score takes the most effective of the two possibilities by looking at efficacy in the biomarker positive and negative groups,
\begin{equation}
\label{eqn:targeted_efficacy}
\mathrm{TargetedEfficacy}(\mathbf{x}_g, j) = \textrm{max}\left( \mathrm{Efficacy}(\mathbf{x}_g, j), \mathrm{Efficacy}(1-\mathbf{x}_g, j) \right) \, .
\end{equation}
Targeted efficacy is distinguished from common measures of drug efficacy, like IC50 and area under the dose-response curve (AUC), in that it is tied directly to the biomarker of interest. Further, rather than summarize entire curves or projecting to the $50\%$ survival point, targeted efficacy aims to capture whether a specific, viable dose is effective for samples with such a biomarker. It also requires a probabilistic modeling approach that can quantify response uncertainty, making it incompatible with traditional point-estimation approaches to dose-response modeling.

\subsection{Discoveries}
\label{subsec:biomarkers:discoveries}
The full list of $p$-values for all $p < 10^{-5}$ biomarkers on all drugs is available in the supplemental material. \cref{tab:biomarkers} highlights three subsets of these biomarkers. 
For each biomarker, we list the following information:
\begin{itemize}
    \item \textbf{Drug}: The drug being tested.
    \item \textbf{Gene}: The gene where the molecular feature is located.
    \item \textbf{Marker}: The molecular feature type representing the biomarker (e.g. mutated or overexpressed).
    \item \textbf{Indicates}: Whether a positive value for this biomarker indicates sensitivity or resistance to treatment with the drug.
    \item \textbf{Score}: The targeted efficacy score.
    \item \textbf{Pos \& Neg}: The number of positive and negative samples in the GDSC dataset.
    \item \textbf{\mbox{$p$}-value}: The associated $p$-value for the biomarker conditional independence test.
\end{itemize}
Any biomarker where the targeted efficacy score was higher than all $10^6$ null samples is reported as being less than $10^{-6}$; additional randomizations could be run to derive a more precise value.

\input{tab_biomarkers}

The first two results, BRAF mutations and Nutlin-3(a), support the qualitative findings by reporting well-known biomarkers with a strong biological basis. Specifically, the drugs identified as having BRAF mutation as a biomarker for sensitivity are all either BRAF or MEK inhibitors. As noted in \cref{subsec:benchmarks:qualitative}, these inhibitors target signaling in the MAPK pathway and BRAF mutations (particularly V600E mutations) are expected to indicate sensitivity to treatment by all the drugs discovered. Similarly, Nutlin-3(a) is found to have TP53 mutations as an indicator of resistance and MDM2 overexpression as an indicator of sensitivity, again matching the expected biological response described in \cref{subsec:benchmarks:qualitative}. Along with these two biomarkers, several other well-known biomarkers of Nutlin-3(a) sensitivity are found, such as overexpression of the apoptosis activator gene BAX~\citep{toshiyuki:reed:1995:bax-p53} and the DNA damage-binding gene DDB2~\citep{shangary:wang:2008:targeting-mdm2-p53}.

The third set of results focuses on fusions of the genes BCR and ABL, which are coded as mutations in our feature set. The detected set of drugs for this biomarker include several drugs (e.g., Nilotinib and Bosutinib) that are well-known to target BCR-ABL fusions~\citep{rix:etal:2007:bcr-abl-inhibitors}. The list also recapitulates the potent inhibition of Axitinib in BCR–ABL cell lines~\citep{pemovska:etal:2015:axitinib-bcr-abl}.
Interestingly, cell lines with BCR-ABL fusions appear to be sensitive to other kinase inhibitors (e.g. NG-25 and NVP-BHG712) whose function on BCR-ABL  has not been previously described. This suggests a broader activity of some of these drugs that have to be validated experimentally.

The discoveries made suggest many potential causal drivers of sensitivity and resistance. However, the Bayesian model does not replace the need for validation experiments. Instead, it enables biologists to guide their experimental planning by suggesting potential drivers of sensitivity and resistance. It is possible that latent confounders, such as DNA methylation, are correlated with the features that we explored. These latent confounders represent a backdoor path~\citep{pearl:2009:causality} through which a non-causal feature may appear to represent a causal link. Making strong causal inference statements about drug sensitivity would require follow-up wet lab experiments that directly intervene on the candidate biomarker.


%% file: tab_biomarkers.tex
\begin{table}[t!]
\centering
\begin{tabular}{llllcccr}
\toprule
Drug & Gene & Marker &  Indicates &  Score &  Pos &  Neg & $p$-value \\
\midrule
       \multicolumn{8}{c}{\textbf{BRAF Mutated}} \\
\midrule
CI-1040      & BRAF & Mutated & Sensitivity & 0.53 & 69 &  752 & $5 \times 10^{-6}$ \\
Dabrafenib   & BRAF & Mutated & Sensitivity & 0.62 & 78 &  813 & $< 1 \times 10^{-6}$ \\
PLX-4720     & BRAF & Mutated & Sensitivity & 0.49 & 70 &  762 & $< 1 \times 10^{-6}$ \\
PLX-4720     & BRAF & Mutated & Sensitivity & 0.53 & 81 &  820 & $5 \times 10^{-6}$ \\
Selumetinib  & BRAF & Mutated & Sensitivity & 0.55 & 79 &  824 & $5 \times 10^{-6}$ \\
\midrule
       \multicolumn{8}{c}{\textbf{Nutlin-3(a)}} \\
\midrule
Nutlin-3a & APOBEC3H & Overexpressed & Sensitivity & 0.54 & 57  & 775 & $3 \times 10^{-6}$ \\
Nutlin-3a & BAX      & Overexpressed & Sensitivity & 0.55 & 133 & 699 & $< 1 \times 10^{-6}$ \\
Nutlin-3a & CYP8B1   & Mutated       & Sensitivity & 0.65 & 5   & 827 & $4 \times 10^{-6}$ \\
Nutlin-3a & DCST2    & Mutated       & Sensitivity & 0.65 & 6   & 826 & $5 \times 10^{-6}$ \\
Nutlin-3a & DDB2     & Overexpressed & Sensitivity & 0.55 & 148 & 684 & $< 1 \times 10^{-6}$ \\
Nutlin-3a & EDA2R    & Overexpressed & Sensitivity & 0.56 & 126 & 706 & $< 1 \times 10^{-6}$ \\
Nutlin-3a & FNIP1    & Mutated       & Sensitivity & 0.65 & 6   & 826 & $4 \times 10^{-6}$ \\
Nutlin-3a & FUT7     & Overexpressed & Sensitivity & 0.60 & 37  & 795 & $3 \times 10^{-6}$ \\
Nutlin-3a & GYG1     & Overexpressed & Sensitivity & 0.52 & 94  & 738 & $1 \times 10^{-6}$ \\
Nutlin-3a & MDM2     & Overexpressed & Sensitivity & 0.54 & 106 & 726 & $1 \times 10^{-6}$ \\
Nutlin-3a & PBXIP1   & Mutated       & Sensitivity & 0.65 & 6   & 826 & $4 \times 10^{-6}$ \\
Nutlin-3a & QSOX1    & Mutated       & Sensitivity & 0.67 & 5   & 827 & $< 1 \times 10^{-6}$ \\
Nutlin-3a & RPS27L   & Overexpressed & Sensitivity & 0.55 & 137 & 695 & $< 1 \times 10^{-6}$ \\
Nutlin-3a & TP53     & Mutated       & Resistance  & 0.48 & 485 & 347 & $< 1 \times 10^{-6}$ \\
Nutlin-3a & ZMAT3    & Overexpressed & Sensitivity & 0.56 & 133 & 699 & $< 1 \times 10^{-6}$ \\
\midrule
       \multicolumn{8}{c}{\textbf{BCR-ABL Fusions}} \\
\midrule
Axitinib     & BCR-ABL &  Mutated & Sensitivity & 0.94 & 6 & 824 & $< 1 \times 10^{-6}$ \\
Bosutinib    & BCR-ABL &  Mutated & Sensitivity & 0.84 & 6 & 824 & $< 1 \times 10^{-6}$ \\
Cabozantinib & BCR-ABL &  Mutated & Sensitivity & 0.73 & 5 & 890 & $< 1 \times 10^{-6}$ \\
CHIR-99021   & BCR-ABL &  Mutated & Sensitivity & 0.54 & 6 & 890 & $4 \times 10^{-6}$ \\
CP466722     & BCR-ABL &  Mutated & Sensitivity & 0.61 & 5 & 894 & $2 \times 10^{-6}$ \\
FR-180204    & BCR-ABL &  Mutated & Sensitivity & 0.56 & 5 & 890 & $< 1 \times 10^{-6}$ \\
HG6-64-1     & BCR-ABL &  Mutated & Sensitivity & 0.80 & 5 & 854 & $< 1 \times 10^{-6}$ \\
JQ1          & BCR-ABL &  Mutated & Sensitivity & 0.74 & 6 & 890 & $4 \times 10^{-6}$ \\
Masitinib    & BCR-ABL &  Mutated & Sensitivity & 0.81 & 5 & 892 & $< 1 \times 10^{-6}$ \\
Methotrexate & BCR-ABL &  Mutated & Sensitivity & 0.70 & 6 & 823 & $4 \times 10^{-6}$ \\
NG-25        & BCR-ABL &  Mutated & Sensitivity & 0.98 & 5 & 890 & $< 1 \times 10^{-6}$ \\
Nilotinib    & BCR-ABL &  Mutated & Sensitivity & 0.95 & 6 & 779 & $< 1 \times 10^{-6}$ \\
NVP-BHG712   & BCR-ABL &  Mutated & Sensitivity & 0.98 & 5 & 890 & $< 1 \times 10^{-6}$ \\
Palbociclib  & BCR-ABL &  Mutated & Sensitivity & 0.57 & 6 & 803 & $7 \times 10^{-6}$ \\
Ponatinib    & BCR-ABL &  Mutated & Sensitivity & 0.80 & 5 & 854 & $< 1 \times 10^{-6}$ \\
QL-XI-92     & BCR-ABL &  Mutated & Sensitivity & 0.71 & 5 & 892 & $< 1 \times 10^{-6}$ \\
Tamoxifen    & BCR-ABL &  Mutated & Sensitivity & 0.61 & 6 & 902 & $5 \times 10^{-6}$ \\
Tivozanib    & BCR-ABL &  Mutated & Sensitivity & 0.80 & 5 & 892 & $< 1 \times 10^{-6}$ \\
TL-1-85      & BCR-ABL &  Mutated & Sensitivity & 0.82 & 5 & 890 & $< 1 \times 10^{-6}$ \\
TL-2-105     & BCR-ABL &  Mutated & Sensitivity & 0.80 & 5 & 894 & $< 1 \times 10^{-6}$ \\
Veliparib    & BCR-ABL &  Mutated & Sensitivity & 0.63 & 6 & 824 & $9 \times 10^{-6}$ \\
\bottomrule
\end{tabular}
\caption{\label{tab:biomarkers} A subset of the biomarkers discovered with $p < 10^{-5}$. The discoveries for BRAF mutations and Nutlin-3(a) mirror the results of the qualitative model evaluation. The BCR-ABL fusion biomarkers recapitulate known results in the literature as well as suggest new avenues for investigation.}
\end{table}

%% file: discussion.tex
\section{Discussion}
\label{sec:discussion}
\subsection{The benefits of modeling uncertainty}
Predictive models for cancer cell line drug response enable science to move at a faster pace. If a predictor can faithfully replicate the outcome of a wet lab experiment, scientists can screen drugs quickly in simulation to find potential therapies worth investigating.
The usefulness of good predictors has led biologists to organize predictive modeling competitions to crowdsource better predictors~\citep{costello:etal:2014:dream-drug-sensitivity} and to build bespoke machine learning models to predict drug response~\citep{menden:etal:2013:dr-nn-and-rf,ammad:etal:2016:dr-kernel-bayes-mf,rampasek:etal:2017:dr-vae}. While these efforts have led to models with improved predictive performance, the target of prediction is a summary statistic derived from preprocessing pipelines such as the approach in \cref{subsec:benchmarks:baseline}.


Compressing each experiment down to a single point estimate of a summary statistic is problematic. At each step in the pipeline, simplifying assumptions remove structure from the model and obfuscate the inherent uncertainty in the measurements, effects, and outcomes. Specifically, (i) averaging the negative controls ignores measurement noise and technical error; (ii) averaging positive controls fails to capture natural variability in cell growth; (iii) a log-linear dose model makes strong assumptions about the effect of different drug concentrations; (iv) the summary statistic reported contains no information about the uncertainty in effect size of a given drug at any specific dose; and (v) biomarker discovery does not account for uncertainty in the summary statistics. 

With these considerations in mind, we proposed a Bayesian approach to modeling dose-response in high-throughput cancer cell line experiments. The Bayesian model addresses the issues with the typical pipeline approach by directly modeling uncertainty at every step: (i-ii) both positive and negative controls are treated as random quantities, with plate-level uncertainty quantification of machine bias and cell growth; (iii) the dose effect is constrained to be monotonic, but is otherwise fully flexible and not limited to the log-linear regime; (iv) the model enables approximate Bayesian posterior inference over the entire dose-response curve for every experiment; and (v) the posterior distributions are used to detect biomarkers through a new probabilistic test statistic.

The Bayesian model outperforms the pipelined approach in benchmarks on the GDSC dataset. However, there is still ample room for improvement. The neural network architecture and training method we used was not explored extensively. Other models or architectures such as those from the existing literature in computational biology~\citep[e.g.][]{rampasek:etal:2017:dr-vae} may yield better performance if combined with a Bayesian model of high-throughput screening experiments. The model could also be improved to better match the data. For instance, some drugs induce total cell death at high doses, and some have no effect at all below a certain concentration. The logistic-MVN model may be a poor fit in these cases since they push the logits to extreme values and consequently dominate the likelihood. We followed a typical processing pipeline to determine mutations, CNVs, and expression levels. These feature pipelines remove uncertainty in the sequencing process that, if modeled directly, may also lead to a better predictive model. We plan to investigate these extensions in future work. Further, since the initial draft of this paper was released, a new version of the GDSC dataset (GDSC2) has been made available. However, the format of the controls and features available has changed in the new dataset; we plan to adapt our GDSC model to GDSC2 in future work.

Finally, cell lines are only simplified, imperfect models of real patient tumors. To be useful in the clinic, a predictive model will need to incorporate real patient data, such as clinical trial and observational patient health records. As a guide for follow-up experiments, cell lines are a useful tool and predictive modeling of cell line drug response can help to better guide the scientist. As more patient data become available, combining our predictive model for cell lines with patient and other (e.g. mice and organoids) data may lead to a better predictor of actual patient drug response.

\subsection{Conclusions}
\label{subsec:introduction:recommendations}
Beyond the specific model and metrics, our experience led to several observations about measurement and modeling in high-throughput cancer drug screening. We conclude by highlighting six key takeaways:
\begin{enumerate}
    \item \textbf{Spatial and temporal batch effects can substantially skew results in high-throughput dose-response experiments.} We showed evidence that experimental controls are systematically contaminated, creating dependency between observations. We detailed a correction approach for legacy data that first detects and discards contaminated observations, then leverages temporal dependencies to compensate for the loss of data.
    \item \textbf{The one-trial-per-dose paradigm is problematic.} Natural variation among cell lines was observed to be substantial in \cref{subsec:model:natural_variation}, with swings as high as $+/-50\%$ of the median control response being common in many experiments. This complicates inference by creating high degrees of uncertainty about any individual experiment.
    \item \textbf{A deep Bayesian dose-response model outperforms the state of the art.} The Bayesian model was shown in \cref{subsec:benchmarks:quantitative} to outperform a state-of-the-art technique for dose-response modeling developed for the GDSC data. The posterior mean has $\approx20\%$ lower mean squared error when predicting a held out dose level.
    \item \textbf{All molecular feature types contain relevant information.} Through an exhaustive feature subset study in \cref{subsec:benchmarks:feature_ablation}, we showed that the model predictions are improved by each of the three subsets of features (mutations, copy number, and expression).
    \item \textbf{Hierarchical probabilistic models can generate biologically-meaningful, nonlinear predictions.} A series of examples in \cref{subsec:benchmarks:qualitative} showed that the model recapitulates known biology. In some cases, this involved nonlinear combinations of features such as needing a high level of expression in one gene and a corresponding wild type of another gene for a specific drug to be effective. This suggests the approach has the potential to be used in exploratory drug discovery experiments where candidate drugs are often tried without a known mechanism of action.
    \item \textbf{Amortized conditional randomization testing provides a scalable, powerful way to detect novel biomarkers.} By imposing a factor model structure on binarized features, we were able to conduct $9.8$M hypothesis tests. The method uncovered both well-studied biomarkers and many new avenues for future research.
\end{enumerate}

%% file: app_plate_layout.tex
\section{Microwell plate layout}
\label{app:plate_layout}
The layout of each microwell plate is shown in \cref{fig:plate_layout}. A single drug is test per orange row in \cref{fig:plate_layout}, with each well in the row receiving a different dose. The dose schedule follows a 2x dilution schedule. Cross-contamination was observed between the negative and positive controls. Substantial intra-plate variability is also seen among positive controls.

\input{fig_plate_layout}

%% file: fig_plate_layout.tex
\begin{figure}[ht]
\centering
\includegraphics[width=0.8\textwidth]{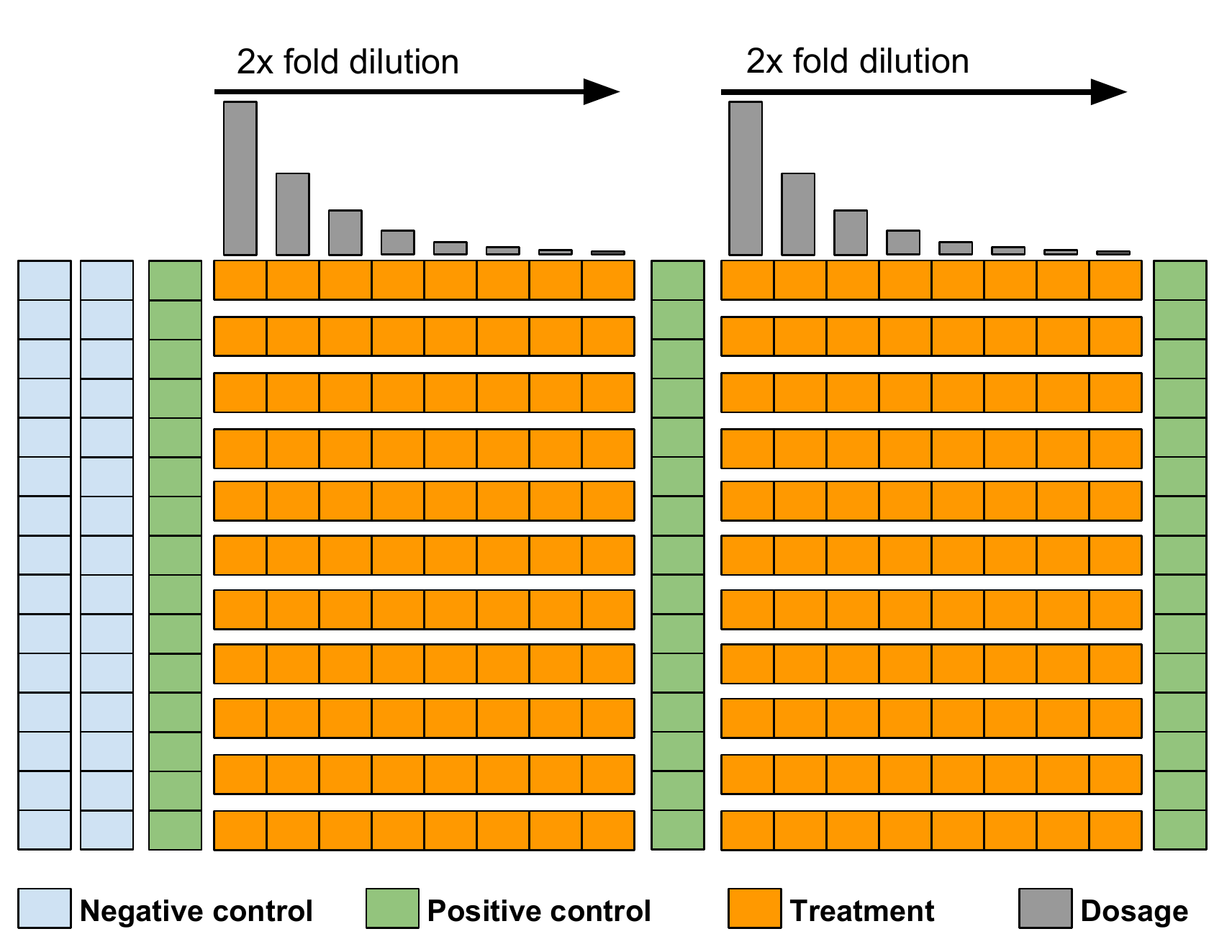}
\caption{\label{fig:plate_layout}Layout of the high-throughput screening plates used in the GDSC experiment for a 9-level dosage schedule. Negative controls (blue) are unpopulated and untreated; positive controls (green) are populated but untreated; treatment wells (orange) are treated with one drug per row, with each well in a row treated at a different concentration.}
\end{figure}

%% file: app_features.tex
\section{Feature preprocessing}
\label{app:features}
\subsection{Mutations}
We consider a gene \textit{mutated} based on the following SNP-level filter:
\begin{itemize}
    \item Silent mutations are ignored.
    \item Insertion, deletion, nonsense, and nonstop mutations are automatically considered mutated.
    \item Missense SNPs are considered mutated if they are present in at least $3$ TCGA \citep{tomczak:etal:2015:tcga} samples.
    \item Splice site mutations are considered mutated if they have an ``IMPACT'' score of high.
\end{itemize}
We also include the two gene fusions that were detected in the GDSC data, EWSR1-X and BCR-ABL. This generates a total of $5822$ mutation features.

\subsection{Copy Number Variations}
Copy number alterations occur when a swath of the genome is lost or amplified. In most cases, this occurs across multiple gene coding regions at once. The result is a highly correlated set of features. We filter this list down to known and potential drivers as follows:
\begin{itemize}
    \item We keep the list of 193 known driver genes identified in the original GDSC analysis.
    \item We add an additional list of 87 potential driver genes identified in a follow up study \citep{zack:etal:2013:cnv-driver-genes}.
    \item We remove genes from the second list if their copy number is constant across all samples or has correlation coefficient $\geq 0.95$ with any gene from the first list.
\end{itemize}
For all cell lines, we consider their median copy number to be the baseline level. Despite the GDSC data coming from human tumor cultured cell lines, the median copy number is actually $3$. Since copy number mostly impacts differential protein expression, relative abundance in a cell is the most important feature here and we thus consider a cell's median copy number to be its baseline, even though in a normal human tissue a copy number of $3$ would be a gain.

\subsection{Expression}
We retain all $17271$ protein-coding genes sequenced in the GDSC dataset. We use the same preprocessed dataset as in the original GDSC analysis.

%% file: app_assumption_checks.tex
\section{Checking drug effect assumptions}
\label{app:modeling_assumptions}
\begin{figure}[t]
\centering
\subfigure[All experiments]{\includegraphics[width=0.49\textwidth]{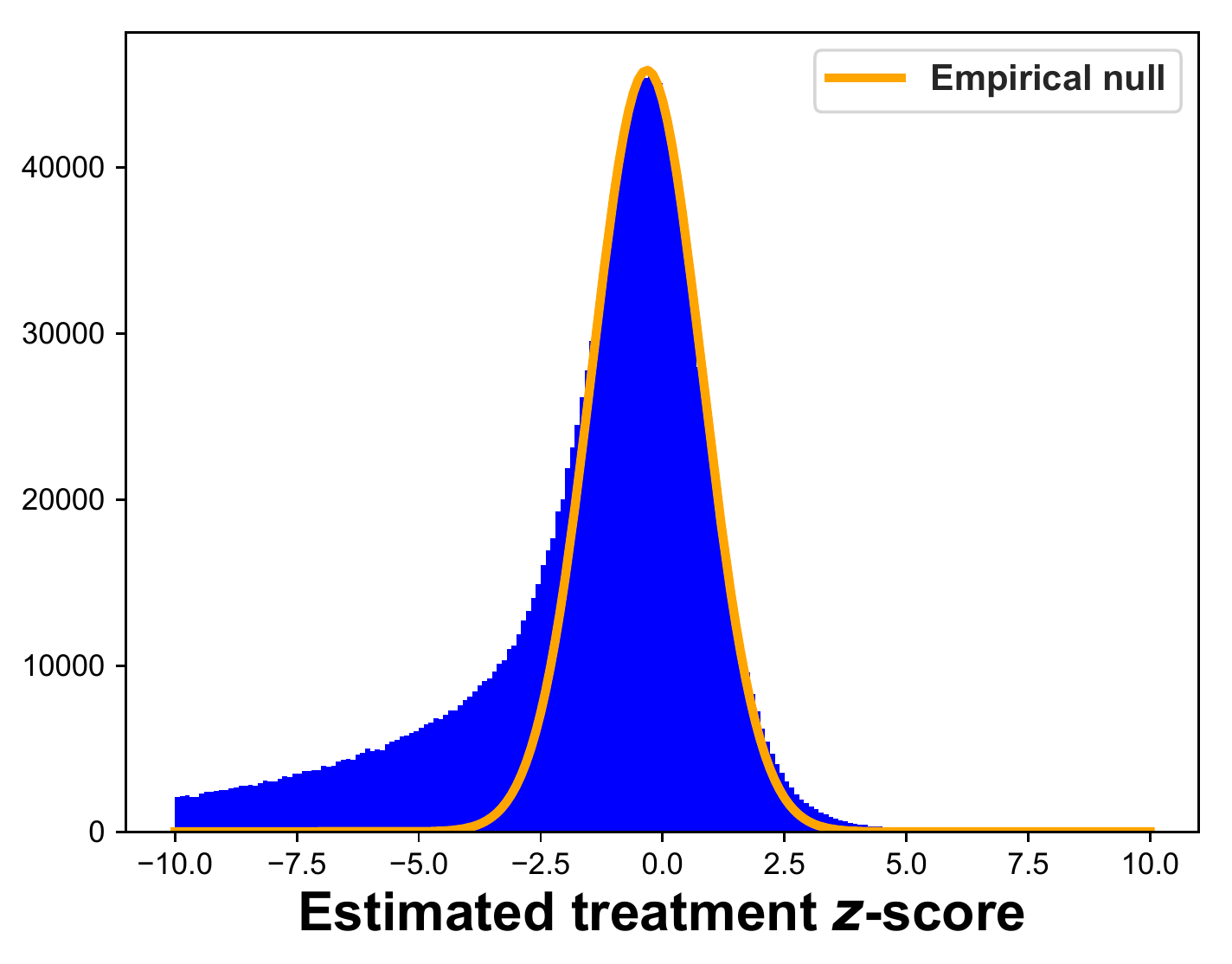}}
\subfigure[Individual drugs]{\includegraphics[width=0.49\textwidth]{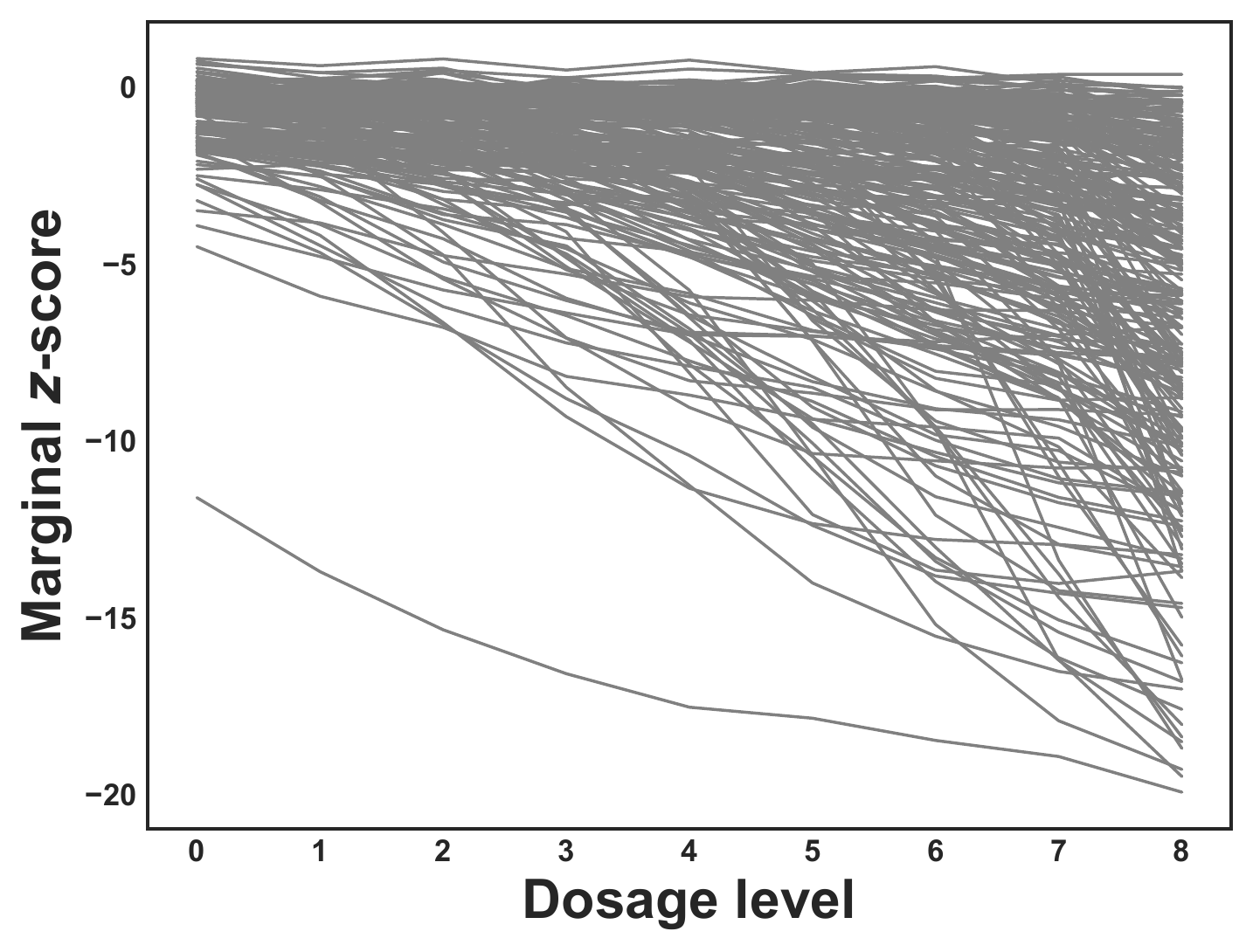}}
\caption{The marginal distribution of approximate $z$-scores from \eqref{eqn:approx_z_scores} across (a) all experiments in the dataset, and (b) individual drugs. In (a), we overlay an empirical estimate of the null distribution; most drugs contain no effect and there appears to be no evidence that any drugs have a positive effect on growth. Similarly, in (b) we observe that all drugs seem to have stronger marginal effects as the dosage level increases, suggesting that the monotone toxicity assumption is also reasonable.}
\label{fig:tau_distribution}
\end{figure}

As a pragmatic check on our assumption that drugs can only hinder growth, we use the estimated control parameters from \cref{sec:model} to calculate approximate $z$-scores for the drug effects,
\begin{equation}
\label{eqn:approx_z_scores}
\widetilde{z}_{ijt} = \frac{y_{ijt} - \hat{a}_{\ell(i,j)} \times \hat{b}_{\ell(i,j)} - \hat{c}_{\ell(i,j)}}{\hat{a}_{\ell(i,j)} \times \hat{b}^2_{\ell(i,j)}} \, .
\end{equation}
Figure~\ref{fig:tau_distribution}a shows the marginal distribution of $\widetilde{\mathbf{z}}$ across all experiments and all dosage levels in the dataset, omitting $z$-scores less than $-10$. The $z$-scores are likely slightly biased due to experimental, technical, and preprocessing error, making the theoretical $\mathcal{N}(0,1)$ null distribution poorly specified; this is a common phenomenon in HTS experiments \citep{efron:2008:twogroups}. We calculate an empirical null using a polynomial approximation to Efron's method \citep{efron:2004}. The empirical null density suggests that the alternative (i.e. non-null treatment effect) distribution has effectively no mass in the positive (i.e. growth-encouraging) region of $z$-scores, supporting our death-only assumption on drug effects.

Figure~\ref{fig:tau_distribution}b presents the same marginal $z$-score check from \eqref{eqn:approx_z_scores}, broken down by individual drugs as a function of dosage level. Every drug appears to have a clear monotone marginal distribution, supporting our montonicity assumption on drug dosage.

%% file: app_cross_contamination.tex
\section{Cross-contamination}
\label{app:cross_contamination}
While bias concerns have been well-investigated in other HTS assays, such concerns are not yet widespread in the cancer research community. Consequently, our high-throughput cancer screening data contains no spatial information about the microwell layout, making it difficult to leverage existing methods of removing spatial bias~\citep[e.g.]{mazoure:etal:2017:hts-spatial-bias}. Instead, the information available is a unique ID for each positive and negative control well (across plates in the same testing site and assay type); no location information about the treatment wells is available outside the generic plate design. Nonetheless, uncovering systematic spatial bias is still crucial in the GDSC data, as the negative and positive control wells are positioned adjacent one another. Therefore, any bleeding over of the positive signal could drastically alter the estimate of the baseline fluorescence.

\begin{figure}[th]
\centering
\includegraphics[width=0.8\textwidth]{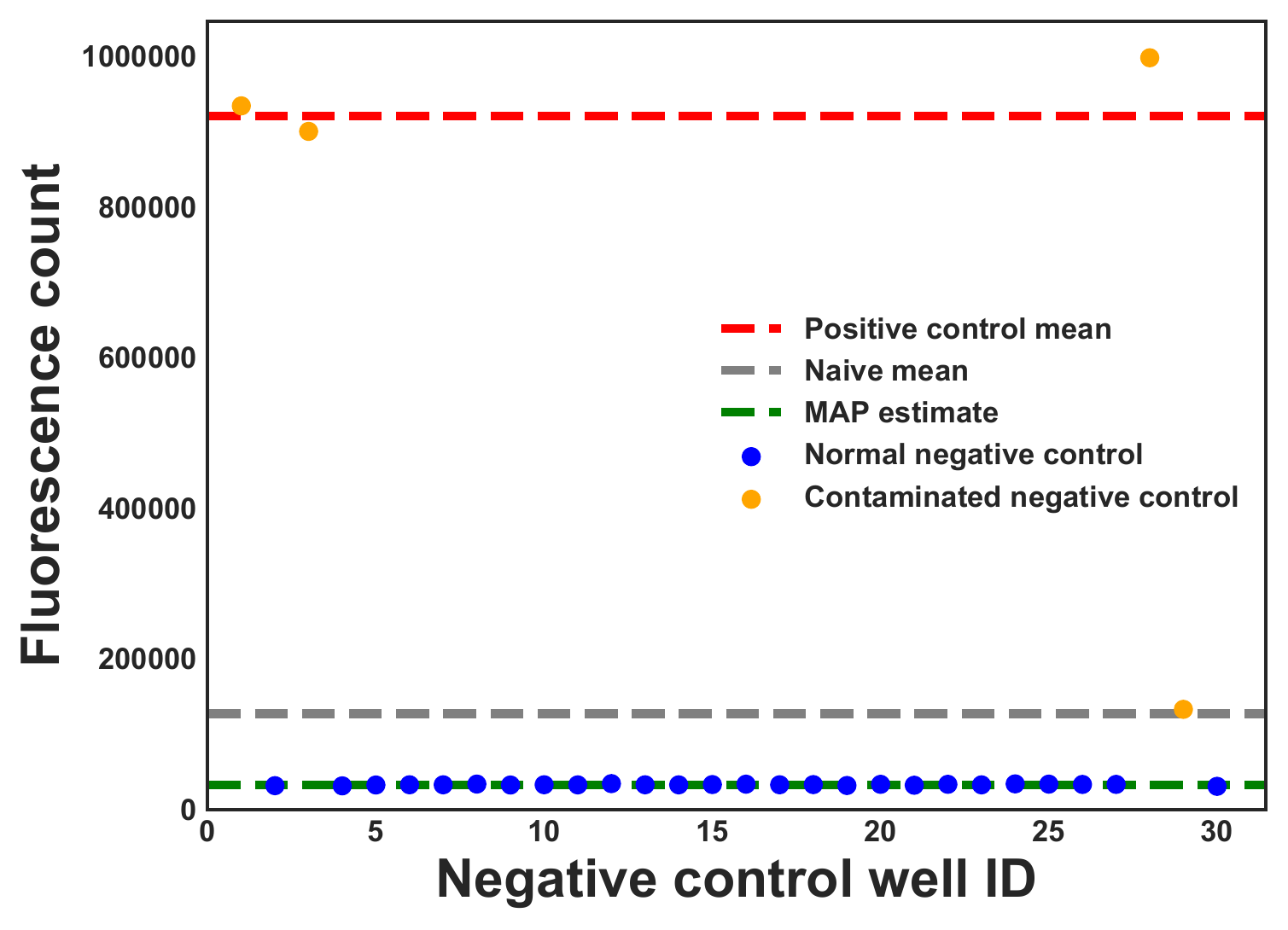}
\caption{Example microwell contamination for one plate in the GDSC data. Each dot is a different negative control well. The dashed red line at the top is the mean of the positive control wells; the middle gray line is the mean of the negative controls; the bottom line is the negative control estimate after our debiasing steps. The improved estimate accounts for an additional $11\%$ of treatment effect for any cells on this plate.}
\label{fig:contamination_example}
\end{figure}

\cref{fig:contamination_example} shows an example of such positive-to-negative control well contamination. In this example, the positive controls are all roughly around $900K$, as indicated by the dashed red mean line at the top. The negative control wells mostly hover around $30K$ but some wells have obviously been contaminated by positive control wells and are wildly different. Estimating the negative control rate as the mean of the wells (dashed gray line) would put the estimate of the bias at approximately $14.5\%$ of the positive control mean. The approximate Bayes procedure we describe below estimates the bias with the contamination effects removed at approximately $3.5\%$. Any treatments estimated from the mean method would therefore bias the drug effects upward, estimating treatment efficacy to be stronger than was likely true.

\begin{figure}[th]
\centering
\includegraphics[width=0.95\textwidth]{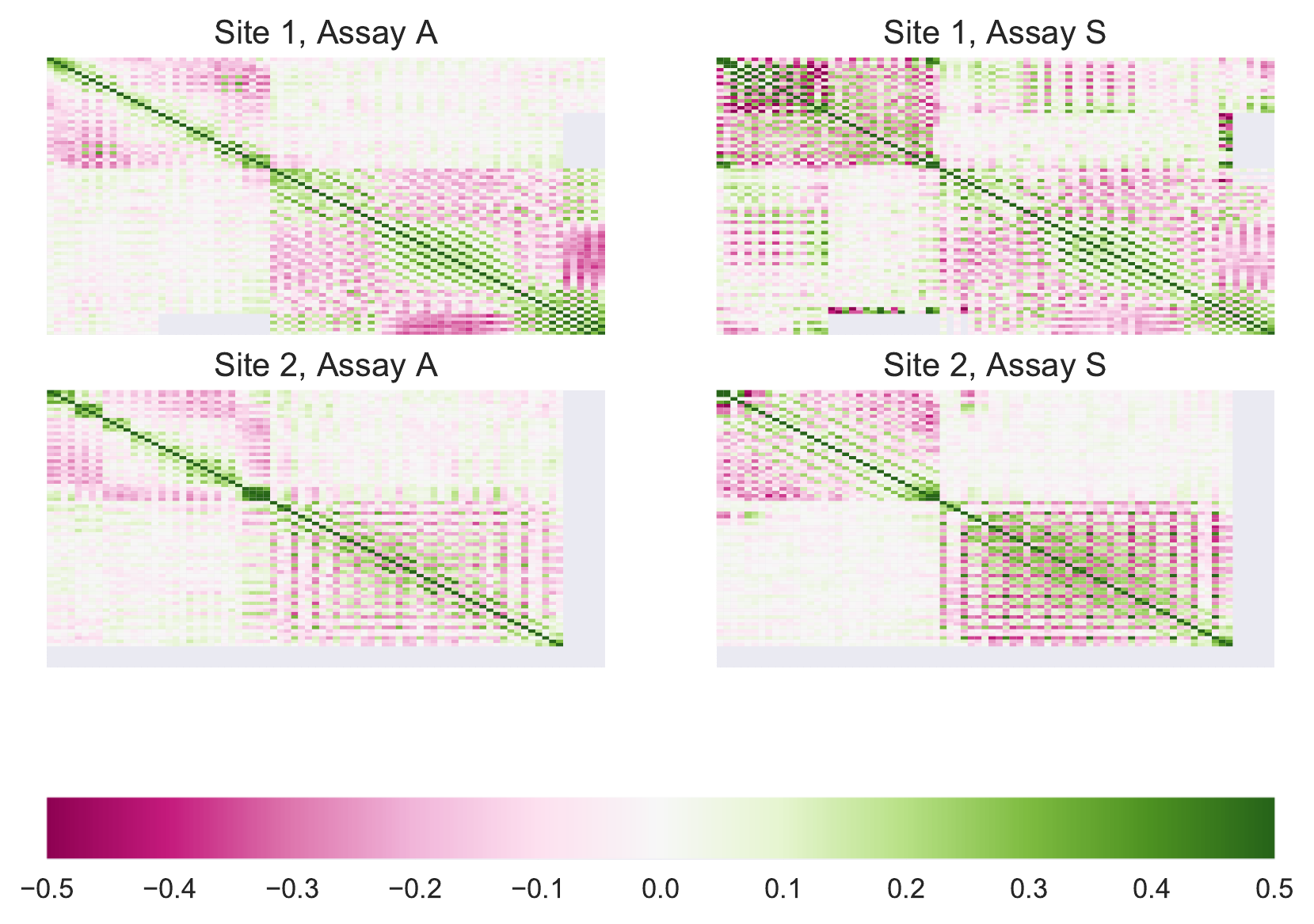}
\caption{Evidence of potential cross-contamination or other systematic biases in the GDSC data. The four figures show control microwell correlation across all plate experiments, stratified by testing site and assay type. The top left corner in each subplot contains the negative control wells; the bottom right contains the positive control wells. Site 1, Assay S in particular shows clear signs of correlation between the two well types, as denoted by the large off-diagonal correlations.}
\label{fig:cross_contamination}
\end{figure}

Without exact layout knowledge, removing the spatial effects is challenging. Fortunately, we have a large number of both negative ($\approx30$) and positive ($\approx40$) controls, along with well-specific IDs; these IDs consistently map to the same microwell location across all plates for the same assay type and at the same testing site. This enables us to determine which negative control wells in each site and assay stratification are being systematically contaminated. \cref{fig:cross_contamination} shows the cross-correlation between all of the control wells in each of the four stratifications; different stratifications have different numbers of control wells. In each subplot, the upper left block represents the correlation structure of the negative controls, the lower right represents the positive controls, and the off-diagonal blocks represent the correlation between negative and positive wells.

Every site shows evidence of at least some contamination, with non-zero entries for many of the off-diagonal blocks. In particular Site 1, Assay S (1S) shows evidence of a large degree of contamination, with the majority of the negative wells being correlated with the positive wells. As a conservative measure, we remove from the dataset any negative wells whose maximum positive well correlation is greater than 0.15 in magnitude. For three of the four testing sites, less than 10 of the negative wells are dropped; for 1S, only 7 negative controls remain.

%% file: app_with_cancer_type.tex
\section{Comparison of model performance with and without cancer type}

We investigated whether using cancer type as an additional feature (i.e. in addition to mutation, CNV, and expression) added predictive power. \cref{tab:curve_prediction_cancer_type} shows the results comparing the two feature sets. Overall, there seems to be no substantial benefit (or cost) to including cancer type as a feature. One possible reason for this is that gene expression and mutation contain enough information to effectively identify the cancer type, leaving it with no conditional mutual information with the drug response.

\input{tab_with_cancer_type}

%% file: tab_with_cancer_type.tex
\begin{table}[th]
\centering
\begin{small}\begin{tabular}{|
p{15mm}|c|c|c|c|c|c|c|c|c|c|}
\hline
& \textbf{(min)} & \multicolumn{7}{|c|}{\textbf{Dose level}} & \textbf{(max)} & \\ \cline{2-10}
\textbf{Model} & \textbf{0} & \textbf{1} & \textbf{2} & \textbf{3} & \textbf{4} & \textbf{5} & \textbf{6} & \textbf{7} & \textbf{8} & \textbf{All} \\ \hline
With Cancer Type           & 1.01 & 1.00 & 1.00 & 1.00 & 1.00 & 1.00 & 1.00 & 1.01 & 1.01 & 1.01 \\ \hline
Without Cancer Type        & 1.00 & 1.00 & 1.00 & 1.00 & 1.00 & 1.00 & 1.00 & 1.00 & 1.00 & 1.00 \\ \hline
\end{tabular}
\end{small}
\caption{\label{tab:curve_prediction_cancer_type} Mean error results when including the cancer type as a feature, in addition to all molecular features, relative to using only the molecular features. Using the cancer type seems to add little-to-no new predictive power to the model.}
\end{table}

%% file: main.bbl
\begin{thebibliography}{40}
\providecommand{\natexlab}[1]{#1}
\providecommand{\url}[1]{\texttt{#1}}
\expandafter\ifx\csname urlstyle\endcsname\relax
  \providecommand{\doi}[1]{doi: #1}\else
  \providecommand{\doi}{doi: \begingroup \urlstyle{rm}\Url}\fi

\bibitem[Ammad-ud din et~al.(2016)Ammad-ud din, Khan, Malani, Murum{\"a}gi,
  Kallioniemi, Aittokallio, and Kaski]{ammad:etal:2016:dr-kernel-bayes-mf}
Muhammad Ammad-ud din, Suleiman~A Khan, Disha Malani, Astrid Murum{\"a}gi, Olli
  Kallioniemi, Tero Aittokallio, and Samuel Kaski.
\newblock Drug response prediction by inferring pathway-response associations
  with kernelized {{B}}ayesian matrix factorization.
\newblock \emph{Bioinformatics}, 32\penalty0 (17):\penalty0 i455--i463, 2016.

\bibitem[Barretina et~al.(2012)Barretina, Caponigro, Stransky, Venkatesan,
  Margolin, Kim, Wilson, Leh{\'a}r, Kryukov, and
  Sonkin]{barretina:etal:2012:ccle}
Jordi Barretina, Giordano Caponigro, Nicolas Stransky, Kavitha Venkatesan,
  Adam~A Margolin, Sungjoon Kim, Christopher~J Wilson, Joseph Leh{\'a}r,
  Gregory~V Kryukov, and Dmitriy Sonkin.
\newblock The cancer cell line encyclopedia enables predictive modelling of
  anticancer drug sensitivity.
\newblock \emph{Nature}, 483\penalty0 (7391):\penalty0 603, 2012.

\bibitem[Candes et~al.(2018)Candes, Fan, Janson, and
  Lv]{candes:etal:2018:panning}
Emmanuel Candes, Yingying Fan, Lucas Janson, and Jinchi Lv.
\newblock Panning for gold: `model-{{X}}' knockoffs for high dimensional
  controlled variable selection.
\newblock \emph{Journal of the Royal Statistical Society: Series B (Statistical
  Methodology)}, 2018.

\bibitem[Cheng et~al.(2015)Cheng, Mitchell, Zehir, Shah, Benayed, Syed,
  Chandramohan, Liu, Won, and Scott]{cheng:etal:2015:msk-impact}
Donavan~T Cheng, Talia~N Mitchell, Ahmet Zehir, Ronak~H Shah, Ryma Benayed,
  Aijazuddin Syed, Raghu Chandramohan, Zhen~Yu Liu, Helen~H Won, and Sasinya~N
  Scott.
\newblock Memorial {{S}}loan {{K}}ettering-integrated mutation profiling of
  actionable cancer targets ({{MSK-IMPACT}}): {{A}} hybridization capture-based
  next-generation sequencing clinical assay for solid tumor molecular oncology.
\newblock \emph{The Journal of molecular diagnostics}, 17\penalty0
  (3):\penalty0 251--264, 2015.

\bibitem[Costello et~al.(2014)Costello, Heiser, Georgii, G{\"o}nen, Menden,
  Wang, Bansal, Hintsanen, Khan, and
  Mpindi]{costello:etal:2014:dream-drug-sensitivity}
James~C Costello, Laura~M Heiser, Elisabeth Georgii, Mehmet G{\"o}nen,
  Michael~P Menden, Nicholas~J Wang, Mukesh Bansal, Petteri Hintsanen,
  Suleiman~A Khan, and John-Patrick Mpindi.
\newblock A community effort to assess and improve drug sensitivity prediction
  algorithms.
\newblock \emph{Nature biotechnology}, 32\penalty0 (12):\penalty0 1202, 2014.

\bibitem[Efron(2004)]{efron:2004}
Bradley Efron.
\newblock Large-scale simultaneous hypothesis testing: the choice of a null
  hypothesis.
\newblock \emph{Journal of the American Statistical Association}, 99\penalty0
  (96--104), 2004.

\bibitem[Efron(2008)]{efron:2008:twogroups}
Bradley Efron.
\newblock Microarrays, empirical {{B}}ayes and the two-groups model (with
  discussion).
\newblock \emph{Statistical Science}, 1\penalty0 (23):\penalty0 1--22, 2008.

\bibitem[Garnett et~al.(2012)Garnett, Edelman, Heidorn, Greenman, Dastur, Lau,
  Greninger, Thompson, Luo, and Soares]{garnett:etal:2012:gdsc}
Mathew~J Garnett, Elena~J Edelman, Sonja~J Heidorn, Chris~D Greenman, Anahita
  Dastur, King~Wai Lau, Patricia Greninger, I~Richard Thompson, Xi~Luo, and
  Jorge Soares.
\newblock Systematic identification of genomic markers of drug sensitivity in
  cancer cells.
\newblock \emph{Nature}, 483\penalty0 (7391):\penalty0 570, 2012.

\bibitem[Haibe-Kains et~al.(2013)Haibe-Kains, El-Hachem, Birkbak, Jin, Beck,
  Aerts, and Quackenbush]{haibe-kains:etal:2013:gdsc-ccle-inconsistency}
Benjamin Haibe-Kains, Nehme El-Hachem, Nicolai~Juul Birkbak, Andrew~C Jin,
  Andrew~H Beck, Hugo~JWL Aerts, and John Quackenbush.
\newblock Inconsistency in large pharmacogenomic studies.
\newblock \emph{Nature}, 504\penalty0 (7480):\penalty0 389--393, 2013.

\bibitem[Haverty et~al.(2016)Haverty, Lin, Tan, Yu, Lam, Lianoglou, Neve,
  Martin, Settleman, and Yauch]{haverty:etal:2016:reproducible}
Peter~M Haverty, Eva Lin, Jenille Tan, Yihong Yu, Billy Lam, Steve Lianoglou,
  Richard~M Neve, Scott Martin, Jeff Settleman, and Robert~L Yauch.
\newblock Reproducible pharmacogenomic profiling of cancer cell line panels.
\newblock \emph{Nature}, 533\penalty0 (7603):\penalty0 333--337, 2016.

\bibitem[Iorio et~al.(2016)Iorio, Knijnenburg, Vis, Bignell, Menden, Schubert,
  Aben, Gon{\c{c}}alves, Barthorpe, Lightfoot, et~al.]{iorio:etal:2016:gdsc}
Francesco Iorio, Theo~A Knijnenburg, Daniel~J Vis, Graham~R Bignell, Michael~P
  Menden, Michael Schubert, Nanne Aben, Emanuel Gon{\c{c}}alves, Syd Barthorpe,
  Howard Lightfoot, et~al.
\newblock A landscape of pharmacogenomic interactions in cancer.
\newblock \emph{Cell}, 166\penalty0 (3):\penalty0 740--754, 2016.

\bibitem[Johnson et~al.(2007)Johnson, Li, and
  Rabinovic]{johnson:etal:2007:batch-effects}
W~Evan Johnson, Cheng Li, and Ariel Rabinovic.
\newblock Adjusting batch effects in microarray expression data using empirical
  {{B}}ayes methods.
\newblock \emph{Biostatistics}, 8\penalty0 (1):\penalty0 118--127, 2007.

\bibitem[Lachmann et~al.(2016)Lachmann, Giorgi, Alvarez, and
  Califano]{lachmann:etal:2016:hts-spatial-bias}
Alexander Lachmann, Federico~M Giorgi, Mariano~J Alvarez, and Andrea Califano.
\newblock Detection and removal of spatial bias in multiwell assays.
\newblock \emph{Bioinformatics}, 32\penalty0 (13):\penalty0 1959--1965, 2016.

\bibitem[Leek et~al.(2010)Leek, Scharpf, Bravo, Simcha, Langmead, Johnson,
  Geman, Baggerly, and Irizarry]{leek:etal:2010:batch-effects}
Jeffrey~T Leek, Robert~B Scharpf, H{\'e}ctor~Corrada Bravo, David Simcha,
  Benjamin Langmead, W~Evan Johnson, Donald Geman, Keith Baggerly, and Rafael~A
  Irizarry.
\newblock Tackling the widespread and critical impact of batch effects in
  high-throughput data.
\newblock \emph{Nature Reviews Genetics}, 11\penalty0 (10):\penalty0 733--739,
  2010.

\bibitem[Lin and Dunson(2014)]{lin:dunson:2014:monotone-gp}
Lizhen Lin and David~B Dunson.
\newblock Bayesian monotone regression using {{G}}aussian process projection.
\newblock \emph{Biometrika}, 101\penalty0 (2):\penalty0 303--317, 2014.

\bibitem[Loken and Gelman(2017)]{loken:gelman:2017:measurement-error}
Eric Loken and Andrew Gelman.
\newblock Measurement error and the replication crisis.
\newblock \emph{Science}, 355\penalty0 (6325):\penalty0 584--585, 2017.

\bibitem[Low-Kam et~al.(2015)Low-Kam, Telesca, Ji, Zhang, Xia, Zink, and
  Nel]{low-kam:etal:2015:bart-dose-response}
Cecile Low-Kam, Donatello Telesca, Zhaoxia Ji, Haiyuan Zhang, Tian Xia,
  Jeffrey~I Zink, and Andre~E Nel.
\newblock A {{B}}ayesian regression tree approach to identify the effect of
  nanoparticles’ properties on toxicity profiles.
\newblock \emph{The Annals of Applied Statistics}, 9\penalty0 (1):\penalty0
  383--401, 2015.

\bibitem[Mazoure et~al.(2017)Mazoure, Nadon, and
  Makarenkov]{mazoure:etal:2017:hts-spatial-bias}
Bogdan Mazoure, Robert Nadon, and Vladimir Makarenkov.
\newblock Identification and correction of spatial bias are essential for
  obtaining quality data in high-throughput screening technologies.
\newblock \emph{Scientific reports}, 7\penalty0 (1):\penalty0 11921, 2017.

\bibitem[Menden et~al.(2013)Menden, Iorio, Garnett, McDermott, Benes,
  Ballester, and Saez-Rodriguez]{menden:etal:2013:dr-nn-and-rf}
Michael~P Menden, Francesco Iorio, Mathew Garnett, Ultan McDermott, Cyril~H
  Benes, Pedro~J Ballester, and Julio Saez-Rodriguez.
\newblock Machine learning prediction of cancer cell sensitivity to drugs based
  on genomic and chemical properties.
\newblock \emph{PLoS one}, 8\penalty0 (4):\penalty0 e61318, 2013.

\bibitem[Muir et~al.(2016)Muir, Li, Lou, Wang, Spakowicz, Salichos, Zhang,
  Weinstock, Isaacs, and Rozowsky]{muir:etal:2016:sequencing-cost}
Paul Muir, Shantao Li, Shaoke Lou, Daifeng Wang, Daniel~J Spakowicz, Leonidas
  Salichos, Jing Zhang, George~M Weinstock, Farren Isaacs, and Joel Rozowsky.
\newblock The real cost of sequencing: {{S}}caling computation to keep pace
  with data generation.
\newblock \emph{Genome biology}, 17\penalty0 (1):\penalty0 53, 2016.

\bibitem[Murray et~al.(2010)Murray, Adams, and
  MacKay]{murray:etal:2010:elliptical-slice-sampling}
Iain Murray, Ryan Adams, and David MacKay.
\newblock Elliptical slice sampling.
\newblock In \emph{Proceedings of the Thirteenth International Conference on
  Artificial Intelligence and Statistics}, pages 541--548, 2010.

\bibitem[Pearl(2009)]{pearl:2009:causality}
Judea Pearl.
\newblock \emph{Causality}.
\newblock Cambridge university press, 2009.

\bibitem[Pemovska et~al.(2015)Pemovska, Johnson, Kontro, Repasky, Chen, Wells,
  Cronin, McTigue, Kallioniemi, Porkka,
  et~al.]{pemovska:etal:2015:axitinib-bcr-abl}
Tea Pemovska, Eric Johnson, Mika Kontro, Gretchen~A Repasky, Jeffrey Chen,
  Peter Wells, Ciar{\'a}n~N Cronin, Michele McTigue, Olli Kallioniemi, Kimmo
  Porkka, et~al.
\newblock Axitinib effectively inhibits bcr-abl1 (t315i) with a distinct
  binding conformation.
\newblock \emph{Nature}, 519\penalty0 (7541):\penalty0 102--105, 2015.

\bibitem[Piovan et~al.(2013)Piovan, Yu, Tosello, Herranz, Ambesi-Impiombato,
  Da~Silva, Sanchez-Martin, Perez-Garcia, Rigo, and
  Castillo]{piovan:etal:2013:akt-inhibition}
Erich Piovan, Jiyang Yu, Valeria Tosello, Daniel Herranz, Alberto
  Ambesi-Impiombato, Ana~Carolina Da~Silva, Marta Sanchez-Martin, Arianne
  Perez-Garcia, Isaura Rigo, and Mireia Castillo.
\newblock Direct reversal of glucocorticoid resistance by {{AKT}} inhibition in
  acute lymphoblastic leukemia.
\newblock \emph{Cancer cell}, 24\penalty0 (6):\penalty0 766--776, 2013.

\bibitem[Polson et~al.(2013)Polson, Scott, and
  Windle]{polson:etal:2013:polya-gamma}
Nicholas~G Polson, James~G Scott, and Jesse Windle.
\newblock Bayesian inference for logistic models using {{P}}{\'o}lya--{{G}}amma
  latent variables.
\newblock \emph{Journal of the American statistical Association}, 108\penalty0
  (504):\penalty0 1339--1349, 2013.

\bibitem[Rampasek et~al.(2017)Rampasek, Hidru, Smirnov, Haibe-Kains, and
  Goldenberg]{rampasek:etal:2017:dr-vae}
Ladislav Rampasek, Daniel Hidru, Petr Smirnov, Benjamin Haibe-Kains, and Anna
  Goldenberg.
\newblock {{Dr. VAE}}: {{D}}rug response variational autoencoder.
\newblock \emph{arXiv preprint arXiv:1706.08203}, 2017.

\bibitem[Rix et~al.(2007)Rix, Hantschel, D{\"u}rnberger, Remsing~Rix,
  Planyavsky, Fernbach, Kaupe, Bennett, Valent, Colinge,
  et~al.]{rix:etal:2007:bcr-abl-inhibitors}
Uwe Rix, Oliver Hantschel, Gerhard D{\"u}rnberger, Lily~L Remsing~Rix, Melanie
  Planyavsky, Nora~V Fernbach, Ines Kaupe, Keiryn~L Bennett, Peter Valent,
  Jacques Colinge, et~al.
\newblock Chemical proteomic profiles of the bcr-abl inhibitors imatinib,
  nilotinib, and dasatinib reveal novel kinase and nonkinase targets.
\newblock \emph{Blood, The Journal of the American Society of Hematology},
  110\penalty0 (12):\penalty0 4055--4063, 2007.

\bibitem[Rodriguez-Barrueco et~al.(2015)Rodriguez-Barrueco, Yu, Saucedo-Cuevas,
  Olivan, Llobet-Navas, Putcha, Castro, Murga-Penas, Collazo-Lorduy, and
  Castillo-Martin]{rodriguez:etal:2015:jak2-inhibition}
Ruth Rodriguez-Barrueco, Jiyang Yu, Laura~P Saucedo-Cuevas, Mireia Olivan,
  David Llobet-Navas, Preeti Putcha, Veronica Castro, Eva~M Murga-Penas, Ana
  Collazo-Lorduy, and Mireia Castillo-Martin.
\newblock Inhibition of the autocrine
  {{IL}}-6--{{JAK2}}--{{STAT3}}--calprotectin axis as targeted therapy for
  {{HR}}-/{{HER}}2+ breast cancers.
\newblock \emph{Genes \& development}, 2015.

\bibitem[Safikhani et~al.(2016)Safikhani, El-Hachem, Quevedo, Smirnov,
  Goldenberg, Birkbak, Mason, Hatzis, Shi, Aerts,
  et~al.]{safikhani:etal:2016:gdsc-ccle-disagreement}
Zhaleh Safikhani, Nehme El-Hachem, Rene Quevedo, Petr Smirnov, Anna Goldenberg,
  Nicolai~Juul Birkbak, Christopher Mason, Christos Hatzis, Leming Shi,
  Hugo~JWL Aerts, et~al.
\newblock Assessment of pharmacogenomic agreement.
\newblock \emph{F1000Research}, 5, 2016.

\bibitem[Shangary and Wang(2008)]{shangary:wang:2008:targeting-mdm2-p53}
Sanjeev Shangary and Shaomeng Wang.
\newblock Targeting the mdm2-p53 interaction for cancer therapy.
\newblock \emph{Clinical Cancer Research}, 14\penalty0 (17):\penalty0
  5318--5324, 2008.

\bibitem[Tibshirani(2014)]{tibshirani:2014:trend-filtering}
Ryan~J Tibshirani.
\newblock Adaptive piecewise polynomial estimation via trend filtering.
\newblock \emph{The Annals of Statistics}, 42\penalty0 (1):\penalty0 285--323,
  2014.

\bibitem[Tieleman and Hinton(2012)]{tieleman:hinton:2012:rmsprop}
Tijmen Tieleman and Geoffrey Hinton.
\newblock Lecture 6.5-rmsprop: {{D}}ivide the gradient by a running average of
  its recent magnitude.
\newblock \emph{COURSERA: Neural networks for machine learning}, 4\penalty0
  (2):\penalty0 26--31, 2012.

\bibitem[Tomczak et~al.(2015)Tomczak, Czerwi{\'n}ska, and
  Wiznerowicz]{tomczak:etal:2015:tcga}
Katarzyna Tomczak, Patrycja Czerwi{\'n}ska, and Maciej Wiznerowicz.
\newblock The cancer genome atlas ({{TCGA}}): an immeasurable source of
  knowledge.
\newblock \emph{Contemporary oncology}, 19\penalty0 (1A):\penalty0 A68, 2015.

\bibitem[Toshiyuki and Reed(1995)]{toshiyuki:reed:1995:bax-p53}
Miyashita Toshiyuki and John~C Reed.
\newblock Tumor suppressor p53 is a direct transcriptional activator of the
  human bax gene.
\newblock \emph{Cell}, 80\penalty0 (2):\penalty0 293--299, 1995.

\bibitem[Vis et~al.(2016)Vis, Bombardelli, Lightfoot, Iorio, Garnett, and
  Wessels]{vis:etal:2016:sanger-dose-response}
Daniel~J Vis, Lorenzo Bombardelli, Howard Lightfoot, Francesco Iorio, Mathew~J
  Garnett, and Lodewyk~FA Wessels.
\newblock Multilevel models improve precision and speed of {{IC50}} estimates.
\newblock \emph{Pharmacogenomics}, 17\penalty0 (7):\penalty0 691--700, 2016.

\bibitem[Wang et~al.(2014)Wang, Smola, and
  Tibshirani]{wang:etal:2014:falling-factorial-basis}
Yu-Xiang Wang, Alex Smola, and Ryan Tibshirani.
\newblock The falling factorial basis and its statistical applications.
\newblock In \emph{International Conference on Machine Learning}, pages
  730--738, 2014.

\bibitem[Wheeler(2019)]{wheeler:2019:bayesian-additive-dose-response}
Matthew~W Wheeler.
\newblock {{B}}ayesian additive adaptive basis tensor product models for
  modeling high dimensional surfaces: an application to high-throughput
  toxicity testing.
\newblock \emph{Biometrics}, 75\penalty0 (1):\penalty0 193--201, 2019.

\bibitem[Yang et~al.(2012)Yang, Soares, Greninger, Edelman, Lightfoot, Forbes,
  Bindal, Beare, Smith, and Thompson]{yang:etal:2012:gdsc}
Wanjuan Yang, Jorge Soares, Patricia Greninger, Elena~J Edelman, Howard
  Lightfoot, Simon Forbes, Nidhi Bindal, Dave Beare, James~A Smith, and
  I~Richard Thompson.
\newblock Genomics of drug sensitivity in cancer ({{GDSC}}): {{A}} resource for
  therapeutic biomarker discovery in cancer cells.
\newblock \emph{Nucleic acids research}, 41\penalty0 (D1):\penalty0 D955--D961,
  2012.

\bibitem[Zack et~al.(2013)Zack, Schumacher, Carter, Cherniack, Saksena, Tabak,
  Lawrence, Zhang, Wala, and Mermel]{zack:etal:2013:cnv-driver-genes}
Travis~I Zack, Steven~E Schumacher, Scott~L Carter, Andrew~D Cherniack, Gordon
  Saksena, Barbara Tabak, Michael~S Lawrence, Cheng-Zhong Zhang, Jeremiah Wala,
  and Craig~H Mermel.
\newblock Pan-cancer patterns of somatic copy number alteration.
\newblock \emph{Nature genetics}, 45\penalty0 (10):\penalty0 1134, 2013.

\bibitem[Zou and Hastie(2005)]{zou:hastie:2005:elastic-net}
Hui Zou and Trevor Hastie.
\newblock Regularization and variable selection via the elastic net.
\newblock \emph{Journal of the Royal Statistical Society: Series B (Statistical
  Methodology)}, 67\penalty0 (2):\penalty0 301--320, 2005.

\end{thebibliography}
